\documentclass[11pt]{article}

\usepackage[final]{acl}

\usepackage{times}
\usepackage{latexsym}

\usepackage[T1]{fontenc}

\usepackage[utf8]{inputenc}

\usepackage{microtype}

\usepackage{inconsolata}

\usepackage{graphicx}

\usepackage{hyperref}
\usepackage{url}
\usepackage{booktabs}

\usepackage{algorithm}
\usepackage{algorithmic}
\usepackage{booktabs}
\usepackage{amsmath}
\usepackage{amsfonts}
\usepackage{subcaption}
\usepackage{wrapfig} 

\usepackage{graphicx}
\usepackage[most]{tcolorbox}
\usepackage{enumitem}
\usepackage{lipsum}
\usepackage{threeparttable}
\usepackage{xcolor}
\usepackage{colortbl}
\usepackage{multirow}
\usepackage{makecell}
\definecolor{mygreen}{rgb}{0,0.5,0}
\definecolor{lightgray}{rgb}{0.9, 0.9, 0.9}
\usepackage{pifont}
\usepackage{amssymb}

\newcommand{\cmark}{\ding{51}} 
\newcommand{\xmark}{\ding{55}} 

\title{FOCAL: Fine-Grained Optimal-Transport-Driven Contrastive Alignment of Language and ECGs with Waveform Enhancement}

\author{
Haitao Li\textsuperscript{1,2}\quad
Che Liu\textsuperscript{3}\quad
Zhengyao Ding\textsuperscript{1}\quad
Ziyi Liu\textsuperscript{4}\quad
Wenqi Shao\textsuperscript{5}\quad
Zhengxing Huang\textsuperscript{1}\thanks{\;Corresponding author.} \\
\\
\textsuperscript{1}Zhejiang University\quad
\textsuperscript{2}Shanghai Innovation Institute\quad
\textsuperscript{3}Imperial College London \\
\textsuperscript{4}Transtek Medical Electronics Co., Ltd\quad
\textsuperscript{5}Shanghai AI Lab \\
\texttt{lihaitao@zju.edu.cn, zhengxinghuang@zju.edu.cn} \\
}

\begin{document}
\maketitle

\begin{abstract}
Electrocardiograms (ECGs) are essential non-invasive tools for diagnosing cardiovascular diseases. While recent multimodal ECG-Report contrastive learning methods have shown promise for zero-shot ECG interpretation, they predominantly rely on global representations, failing to capture the fine-grained relationship between localized waveform patches and specific pathological tags. This limitation is further exacerbated by the fact that nearly 55\% of standard clinical reports (e.g., in MIMIC-ECG) lack explicit waveform descriptions. In this paper, we propose FOCAL, a novel framework that achieves precise, fine-grained alignment between localized ECG segments and individual report tags via Optimal Transport. Furthermore, because fine-grained alignment at the tag level exacerbates the false negative problem among reports sharing common diagnoses, we introduce a semantic similarity matrix to guide the contrastive objective and correct misalignments. To address the scarcity of detailed waveform text, we introduce a coarse-to-fine enrichment pipeline that leverages Large Language Models (LLMs) to recover missing semantics, utilizing a coarse model verification step to rigorously filter out hallucinations. Extensive experiments across six datasets demonstrate that FOCAL establishes new state-of-the-art performance in zero-shot prediction and linear probing. Our code and data are available at: \url{https://github.com/CuCl-2/FOCAL}.
\end{abstract}


\section{Introduction}
Electrocardiograms (ECGs) are essential non-invasive tools for detecting cardiac rhythm disorders in clinical practice \citep{sahoo2020machine,rath2021heart,ayano2022interpretable,li2026anyecg,ding2025ai}. Recently, self-supervised learning (SSL) has emerged as a promising paradigm for ECG representation learning, alleviating the reliance on large-scale annotated data and expert knowledge. Existing ECG SSL approaches can be broadly categorized into comparative methods \citep{chen2020simple,chen2021empirical,wang2023adversarial,eldele2021time} and generative methods \citep{zhang2022maefe,hu2023spatiotemporal,na2024guiding,zhang2022self}. However, most of these methods still struggle with unseen classes in zero-shot scenarios.


Recent studies leverage multimodal ECG-report alignment to enable zero-shot prediction. While foundational works \citep{li2024frozen,lalam2023ecg} directly map paired ECG signals to clinical texts, subsequent methods introduce sophisticated enhancements: MERL \citep{liu2024zero} incorporates uni-modal alignment alongside LLM descriptive prompts, and ESI \citep{yu2024ecg} integrates RAG \citep{ni2025towards,gao2023retrieval} pipelines. Furthermore, MELP \citep{wang2025token} captures hierarchical ECG structures via multi-scale supervision, while D-BETA \citep{hung2025boosting} combines generative masked auto-encoders with discriminative contrastive learning.

Despite these advances, current methods predominantly focus on aligning global ECG representations with entire reports. This macroscopic approach neglects the fine-grained, localized relationship between specific pathological tags and their exact waveform manifestations. Achieving this fine-grained alignment presents three critical challenges. First, existing architectures rely on global embeddings that fail to capture patch-level ECG features and tag-specific textual semantics. This limitation is further compounded by the lack of explicit annotations mapping individual report tags to their corresponding ECG segments, making direct supervision for such localized alignment impossible. Second, performing alignment at the individual tag level inevitably exacerbates the false negative problem during contrastive learning, as different reports frequently share common tags. Finally, nearly 55\% of ECG reports in the MIMIC-ECG dataset lack explicit waveform descriptions. Recovering these missing features using LLMs is complicated by the inherent hallucination problem and the complex, non-bijective relationship between waveform manifestations and diagnostic outcomes \citep{jin2018screening}. 

To address these challenges, we propose \textbf{FOCAL} (\textbf{F}ine-Grained \textbf{O}ptimal-Transport-Driven \textbf{C}ontrastive \textbf{A}lignment of \textbf{L}anguage and ECGs). First, we formulate the alignment between independent report tags and ECG patches as an Optimal Transport (OT) problem. This strict, parameter-free routing ensures that each pathological tag is forcefully grounded to its most distinct localized waveform features. Second, we introduce a semantic similarity matrix to dynamically mitigate the false negative collisions that occur when processing flattened, tag-level contrastive objectives. Finally, we design a coarse-to-fine LLM enrichment pipeline to recover missing waveform descriptions. Because LLM outputs are susceptible to hallucinations and complex non-bijective mappings, we employ a coarse FOCAL model to rigorously validate the generated waveform features before incorporating them into the pre-training corpus. 

We validate our proposed FOCAL framework on six downstream ECG multi-label classification datasets. The results demonstrate that FOCAL establishes new state-of-the-art performance in zero-shot prediction, linear probing, and cross-domain transfer learning. Overall, our main contributions are as follows:
\begin{itemize}
    \item We propose FOCAL, an Optimal Transport-driven architecture that achieves precise, fine-grained alignment between specific report tags and localized ECG patches.
    \item We present a semantic similarity matrix to guide contrastive learning, actively identifying and mitigating the false negative misalignments exacerbated by tag-level alignment.
    \item We introduce a coarse-to-fine training process that leverages LLMs to recover missing waveform features and validates them with a coarse model, effectively filtering out hallucinations.
    \item Extensive experiments across six datasets demonstrate that FOCAL achieves state-of-the-art performance.
\end{itemize}

\section{Related Work}
\noindent \textbf{ECG-Report Contrastive Learning}
Recent advancements in multimodal contrastive learning \citep{radford2021learning} have spurred significant efforts in aligning ECG signals with clinical texts. Foundational approaches \citep{li2024frozen,liu2024etp,lalam2023ecg} learn representations by directly pulling paired ECGs and reports together while pushing unpaired ones apart. Building on this, MERL \citep{liu2024zero} introduces uni-modal alignment and employs the CKEPE pipeline at inference to generate more descriptive prompts via LLMs. ESI \citep{yu2024ecg} utilizes a retrieval-augmented generation (RAG) pipeline during training, integrating external medical knowledge to formulate more detailed descriptions. To capture the hierarchical nature of ECG signals, MELP \citep{wang2025token} introduces multi-scale cross-modal supervision at the token, beat, and rhythm levels. Concurrently, D-BETA \citep{hung2025boosting} proposes a hybrid framework integrating generative masked auto-encoders with boosted discriminative contrastive learning.

Despite these advances, most of the existing methods perform contrastive learning at the whole-ECG and report level, and overlook the absence of fine-grained waveform features in ECG reports.

\noindent \textbf{Fine-Grained Contrastive Learning}
Recent studies in medical imaging have demonstrated the benefits of fine-grained alignment. Methods such as MedFILIP \citep{liang2025medfilip}, fVLM \citep{shui2025large} incorporate entity-level or region-level supervision by leveraging structured information extracted from reports or by segmenting images into anatomical regions. However, the application of fine-grained contrastive learning in the ECG domain remains largely unexplored.

\noindent \textbf{False Negatives in Contrastive Learning}
Traditional contrastive learning \citep{radford2021learning} assumes all unpaired samples are strict negatives. This fails in the ECG domain, where frequent normal baselines and common diseases create inherent false negatives. Prior solutions involve adding regularization terms \citep{lavoie2024modeling,jiang2023vision,li2023integrating} or inter-report similarity matrices \citep{sun2023learning,wang2022medclip,kim2025falcon}. In our framework, fine-grained alignment at the individual tag level significantly exacerbates this false negative problem since distinct reports frequently share identical diagnostic tags.

\begin{figure*}[t]
    \centering
    \includegraphics[width=1.0\linewidth]{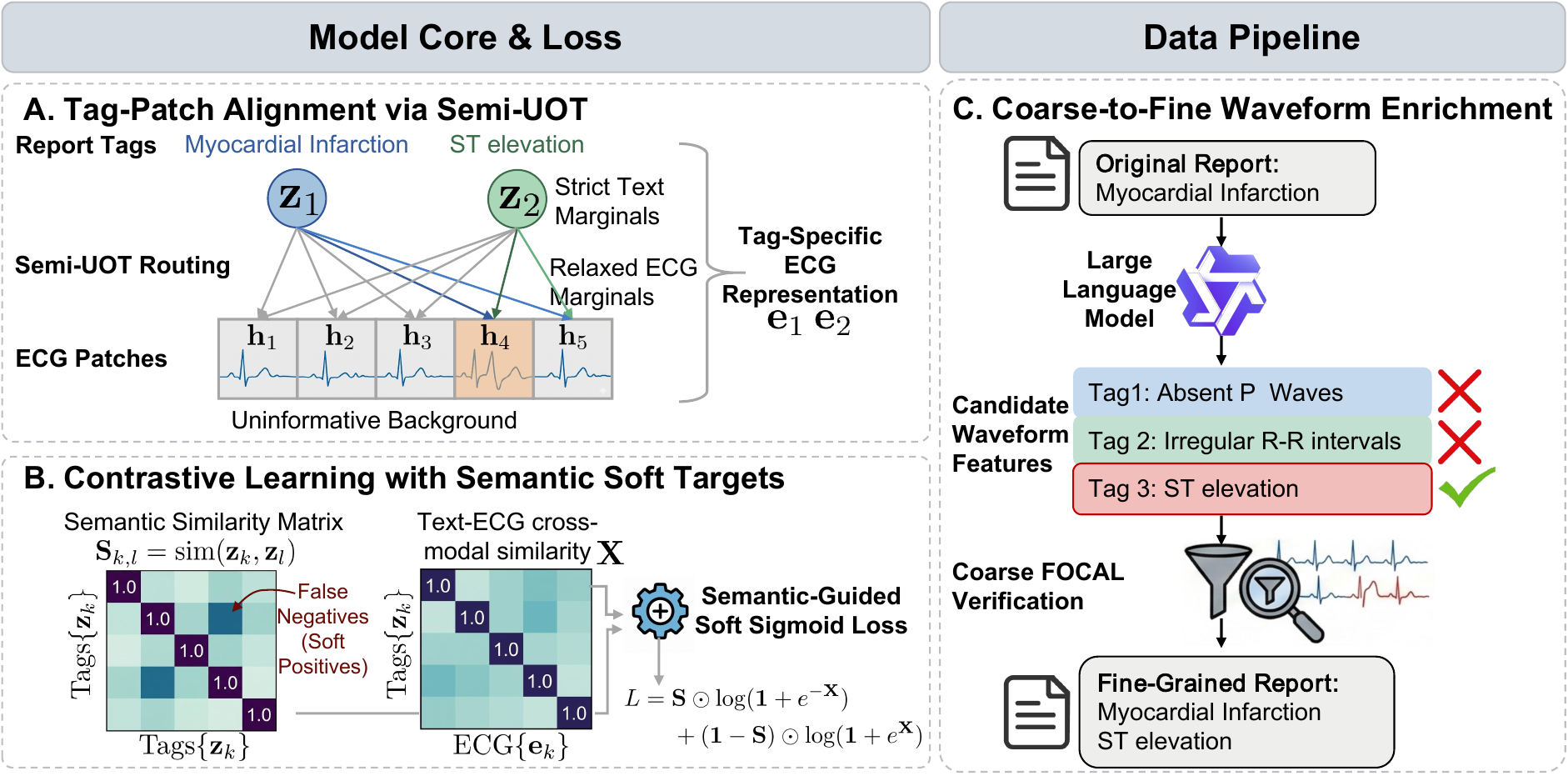}
    \caption{The overall framework of \textbf{FOCAL}. \textbf{(A) Tag-Patch Alignment via Semi-UOT:} Semi-Unbalanced Optimal Transport routes text tags to ECG patches, grounding all clinical semantics while dynamically ignoring uninformative background signals. \textbf{(B) Contrastive Learning with Semantic Soft Targets:} Integrating a text-derived semantic similarity matrix $\mathbf{S}$ into a soft sigmoid objective mitigates false negative bias by treating semantic identical tags across reports as soft positive targets. \textbf{(C) Coarse-to-Fine Waveform Enrichment:} To recover omitted clinical waveform features, an LLM generates candidates from the original report, which are rigorously verified by a Coarse FOCAL model to filter hallucinations and yield enriched reports.}
    \label{fig:fig1}
\end{figure*}



\section{Method}
\subsection{Fine-Grained Alignment Architecture}
In this section, we describe how we achieve fine-grained alignment between specific ECG segments and tags in clinical reports. Given an ECG-report dataset $\mathcal{D} = \{ (\mathbf{X}_i, \mathcal{R}_i) \}$, each report $\mathcal{R}_i$ consists of multiple clinically meaningful tags, such as \textit{arrhythmia}, \textit{myocardial infarction}, \textit{atrial fibrillation}, etc. For each tag, our objective is to identify and align the corresponding ECG segment(s) that reflect this clinical finding.

\paragraph{Spatial-Temporal ECG Encoding.}
To capture both spatial relationships and temporal dynamics, we employ a Spatial-Temporal ViT encoder. Given a 12-lead ECG $\mathbf{X}_i \in \mathbb{R}^{L \times T}$ ($L = 12$, $T$ is the signal length), we divide each lead into $N_{time}$ non-overlapping patches. This yields a total of $N = L \times N_{time}$ patches, which are flattened into a 1D sequence. Each patch $\mathbf{x}_i^p$ ($p = 1, \dots, N$) is linearly projected to $d$ dimensions. To preserve coordinate information, we add learnable lead and temporal embeddings to each projected patch:
\begin{equation}
    \mathbf{u}_i^p = \mathrm{Linear}(\mathbf{x}_i^p) + \mathbf{e}_{lead}^p + \mathbf{e}_{time}^p
\end{equation}
where $\mathbf{e}_{lead}^p$ and $\mathbf{e}_{time}^p$ denote the specific lead and temporal embeddings corresponding to the $p$-th patch. The token sequence $\mathbf{U}_i = [\mathbf{u}_i^1, \dots, \mathbf{u}_i^N]$ is then processed by the ECG encoder to yield the final fine-grained patch embeddings $\mathbf{H}_i = [\mathbf{h}_i^1, \dots, \mathbf{h}_i^N]$:
\begin{equation}
    \mathbf{H}_i = \mathrm{ECGEncoder}(\mathbf{U}_i) \in \mathbb{R}^{N \times d}
\end{equation}

\paragraph{Report Tag-Level Encoding.}
Since ECG reports are highly structured, we can simply use commas to split the tags. We denote the set of tags in the $i$-th report as $\mathcal{T}_i = \{ t_i^j \}_{j=1}^{m_i}$, where $t_i^j$ represents the $j$-th tag in report $i$, and $m_i$ is the number of tags in $\mathcal{R}_i$. Each tag $t_i^j$ is then independently encoded via a text encoder to obtain its embedding representation. This yields the tag-level representation matrix for report $i$:
\begin{equation}
    \mathbf{Z}_i = [\mathbf{z}_i^1, \dots, \mathbf{z}_i^{m_i}] \in \mathbb{R}^{m_i \times d}
\end{equation}
\begin{equation}
    \mathbf{z}_i^j = \mathrm{TextEncoder}(t_i^j), \quad j = 1, \dots, m_i
\end{equation}
Notably, both the ECG patch embeddings and tag embeddings are projected into the same latent space with dimension $d$. For clarity and conciseness, we omit explicit projection formulas in the equations.

\paragraph{Tag-Specific ECG Representation via Semi-Unbalanced Optimal Transport.}
With the fine-grained ECG patch embeddings and report tag embeddings, our goal is to align them accurately. Standard Optimal Transport (OT) typically formulates such alignment by imposing strict, uniform marginal constraints on both modalities. However, this completely balanced formulation is ill-suited for the clinical reality of ECG-report pairs due to a structural asymmetry. On the textual side, every semantic tag extracted from the report contains critical diagnostic information and must be strictly grounded to the waveform. On the signal side, pathological manifestations in an ECG are highly localized and temporally sparse. Consequently, many background patches (e.g., normal baselines or regular heartbeats) carry no relevant diagnostic information. Forcing a strict uniform constraint on the ECG patches would erroneously compel the model to distribute the semantic mass of pathological tags onto these uninformative segments. 

To resolve this asymmetry, we formulate the tag-patch alignment as a Semi-Unbalanced Optimal Transport (Semi-UOT) problem. Specifically, we treat the $m_i$ tag embeddings $\mathbf{Z}_i$ and $N$ ECG patch embeddings $\mathbf{H}_i$ as two discrete distributions. We first define a cost matrix $\mathbf{C} \in \mathbb{R}^{m_i \times N}$, where each element $C_{j,p}$ represents the cosine distance between the $j$-th tag and the $p$-th ECG patch:
\begin{equation}
    C_{j,p} = 1 - \frac{(\mathbf{z}_i^j)^\top \mathbf{h}_i^p}{\|\mathbf{z}_i^j\|_2 \|\mathbf{h}_i^p\|_2}
\end{equation}

Our objective is to find a transport plan $\mathbf{T} \in \mathbb{R}_{+}^{m_i \times N}$ that minimizes the overall transport cost. To accommodate the structural asymmetry, we strictly ensure every tag is grounded, but relax the constraint on ECG patches. To guarantee differentiability and computational efficiency, we introduce an entropic regularization term $\mathcal{H}(\mathbf{T})$ and a Kullback-Leibler (KL) divergence penalty, solving the following objective:
\begin{equation}
\begin{aligned}
    \mathbf{T}^* &= \arg\min_{\mathbf{T} \mathbf{1}_N = \mathbf{a}} \Bigg\{ \sum_{j=1}^{m_i} \sum_{p=1}^{N} T_{j,p} C_{j,p} \\
    &\quad - \epsilon \mathcal{H}(\mathbf{T}) + \tau \text{KL}(\mathbf{T}^\top \mathbf{1}_{m_i} \| \mathbf{b}) \Bigg\}
\end{aligned}
\end{equation}
where $\mathbf{a} = \frac{1}{m_i}\mathbf{1}_{m_i}$ and $\mathbf{b} = \frac{1}{N}\mathbf{1}_N$ represent the prior marginal distributions. The strict equality constraint $\mathbf{T} \mathbf{1}_N = \mathbf{a}$ ensures all clinical tags are fully assigned (the ``balanced'' aspect), while the KL divergence softens the marginal constraint on the ECG patches (the ``unbalanced'' aspect). The hyperparameter $\tau$ controls this relaxation: a smaller $\tau$ allows the network to confidently assign near-zero mass to irrelevant ECG patches. This objective can be efficiently solved using the unbalanced Sinkhorn-Knopp algorithm.

Finally, to aggregate information from the most relevant ECG segments for each tag, we normalize the optimal transport plan such that each row sums to $1$ (i.e., $\hat{T}_{j,p} = m_i \cdot T^*_{j,p}$). The Semi-UOT-routed, tag-specific ECG representation is then computed as a transport-weighted sum:
\begin{equation}
    \mathbf{e}_i^j = \sum_{p=1}^{N} \hat{T}_{j,p} \cdot \mathbf{h}_i^p
\end{equation}
By leveraging Semi-UOT, each tag forcefully captures its most distinct local ECG features while naturally filtering out noisy or irrelevant baselines. Importantly, because the iterative unbalanced Sinkhorn algorithm consists entirely of differentiable matrix operations, this module acts as a parameter-free layer. It can be seamlessly optimized end-to-end with the subsequent contrastive loss, allowing gradients to flow directly back to refine both the ECG and text encoders.

\paragraph{Fine-Grained Contrastive Learning with Semantic Similarity Matrix.}
To maximize data efficiency and construct a comprehensive negative sample pool, we leverage all available tags within a batch. For a batch of $B$ reports, let $K = \sum_{i=1}^B m_i$ denote the total number of tags. We flatten these tags and their corresponding tag-specific ECG representations into two aligned sequences: $\{ \mathbf{z}_k \}_{k=1}^K$ and $\{ \mathbf{e}_k \}_{k=1}^K$.

Standard contrastive objectives, such as the pairwise Sigmoid loss (SigLIP) \citep{zhai2023sigmoid}, treat all non-paired samples as strict negatives. This hard-negative assumption is problematic in fine-grained medical alignment, as different reports frequently share identical pathological tags. Forcing the model to push apart these semantically equivalent but non-paired samples exacerbates the false negative problem. 

To address this, we introduce a semantic similarity matrix $\mathbf{S} \in [0, 1]^{K \times K}$ computed from the text tag embeddings, where $\mathbf{S}_{k,l} = \text{sim}(\mathbf{z}_k, \mathbf{z}_l)$. Instead of using hard binary indicators, we replace them with this continuous semantic prior, formulating a semantic-guided soft sigmoid Loss:
\begin{equation}
\begin{split}
    L &= \frac{1}{K^2} \sum_{k=1}^{K} \sum_{l=1}^{K} \Big[ \mathbf{S}_{k,l} \log\left(1 + \exp(-x_{k,l})\right) \\
    &\quad + (1 - \mathbf{S}_{k,l}) \log\left(1 + \exp(x_{k,l})\right) \Big]
\end{split}
\end{equation}
where $x_{k,l} = t \cdot \text{sim}(\mathbf{e}_k, \mathbf{z}_l)$ is the temperature-scaled cosine similarity between the $k$-th ECG representation and the $l$-th text tag, and $t$ is a learnable temperature parameter.

By integrating $\mathbf{S}$ directly into the binary cross-entropy formulation as soft targets, our model naturally mitigates the false negative bias. Semantically identical tags from different reports ($\mathbf{S}_{k,l} \approx 1$) act as positive soft-targets, elegantly eliminating the erroneous repulsive force at the gradient level. Through this unified process, pathological semantics are explicitly grounded to their most relevant ECG waveform manifestations.

\subsection{Coarse-to-Fine Training Process}
Fine-grained alignment relies on reports that include detailed waveform features. However, we observe that due to clinical habits, these intermediate diagnostic observations are frequently omitted from recorded reports. To recover this missing information, we query LLMs with the question, \textit{`What waveform features are most likely to be present in electrocardiograms with these symptoms?'} to identify potential overlooked characteristics. While this generates potential features, LLM outputs remain unreliable for two primary reasons. First, ECG waveform features and diagnoses are not in a one-to-one correspondence \citep{jin2018screening}. A single disease may present different waveform characteristics across individuals, meaning doctors can infer a diagnosis from waveform features, but the reverse is not strictly possible. Second, the LLM's output is inherently unstable due to hallucination issues \citep{huang2023survey, gunay2024accuracy}.

To address these reliability challenges, we propose a novel coarse-to-fine training process, illustrated in Figure \ref{fig:fig1}, consisting of three steps. We first train a coarse FOCAL model using contrastive learning on the original ECG-report pairs. Next, we validate the LLM-generated waveform features by computing their similarity to the actual ECG signal using this coarse model, selecting only high-confidence candidates for augmentation. Finally, these validated features are incorporated into the original report, and we continue training the coarse model with this augmented data to obtain the final FOCAL model.

\begin{table*}[t]
  \centering
  \small
  \caption{Results of zero-shot classification.}
  \resizebox{0.95\textwidth}{!}{
  \begin{tabular}{lccccccc}
    \toprule
    \textbf{Macro AUC} & \textbf{PTBXL-Super} & \textbf{PTBXL-Sub} & \textbf{PTBXL-Form} & \textbf{PTBXL-Rhythm} & \textbf{CPSC2018} & \textbf{CSN} & \textbf{Average} \\
    \midrule
    MERL \citep{liu2024zero}  & 74.2 & 75.7 & 65.9 & 78.5 & 82.8 & 74.4 & 75.3 \\
    MELP \citep{wang2025token} & 76.2 & 81.2 & 69.1 & 85.4 & 84.2 & 77.6 & 79.0 \\
    D-BETA \citep{hung2025boosting} & 76.2 & 75.9 & 66.1 & 88.6 & 80.1 & 76.3 & 77.1 \\
    \textbf{FOCAL(Ours)} & \textbf{77.7} & \textbf{83.5} & \textbf{70.3} & \textbf{90.8} & \textbf{85.0} & \textbf{79.2} & \textbf{81.1} \\
    \bottomrule
  \end{tabular}
  }
\label{table:zero shot}
\end{table*}

\begin{table*}[ht]
    \centering
    \caption{Results of Linear Evaluation. The best and second-best results are highlighted in bold and \underline{underlined}.}
    \label{tab:linear_probe}
    \resizebox{0.95\textwidth}{!}{
    \begin{tabular}{lccc|ccc|ccc|ccc|ccc|ccc}
        \toprule
      \multirow{2}{*}{\textbf{Method}} 
      & \multicolumn{3}{c}{\textbf{PTBXL-Super}} 
      & \multicolumn{3}{c}{\textbf{PTBXL-Sub}} 
      & \multicolumn{3}{c}{\textbf{PTBXL-Form}} 
      & \multicolumn{3}{c}{\textbf{PTBXL-Rhythm}} 
      & \multicolumn{3}{c}{\textbf{CPSC2018}} 
      & \multicolumn{3}{c}{\textbf{CSN}} \\
      
      & \textbf{1\%} & \textbf{10\%} & \textbf{100\%}
      & \textbf{1\%} & \textbf{10\%} & \textbf{100\%}
      & \textbf{1\%} & \textbf{10\%} & \textbf{100\%}
      & \textbf{1\%} & \textbf{10\%} & \textbf{100\%}
      & \textbf{1\%} & \textbf{10\%} & \textbf{100\%}
      & \textbf{1\%} & \textbf{10\%} & \textbf{100\%} \\
        \midrule
        Random Init & 70.45 & 77.09 & 81.61 & 55.82 & 67.60 & 77.91 & 55.82 & 62.54 & 73.00 & 46.26 & 62.36 & 79.29 & 54.96 & 71.47 & 78.33 & 47.22 & 63.17 & 73.13 \\
        SimCLR & 63.41 & 69.77 & 73.53 & 60.84 & 68.27 & 73.39 & 54.98 & 56.97 & 62.52 & 51.41 & 69.44 & 77.73 & 59.78 & 68.52 & 76.54 & 59.02 & 67.26 & 73.20 \\
        BYOL & 71.70 & 73.83 & 76.45 & 57.16 & 67.44 & 71.64 & 48.73 & 61.63 & 70.82 & 41.99 & 74.40 & 77.17 & 60.88 & 74.42 & 78.75 & 54.20 & 71.92 & 74.69 \\
        BarlowTwins & 72.87 & 75.96 & 78.41 & 62.57 & 70.84 & 74.34 & 52.12 & 60.39 & 66.14 & 50.12 & 73.54 & 77.62 & 55.12 & 72.75 & 78.39 & 60.72 & 71.64 & 77.43 \\
        MoCo-v3 & 73.19 & 76.65 & 78.26 & 55.88 & 69.21 & 76.69 & 50.32 & 63.71 & 71.31 & 51.38 & 71.66 & 74.33 & 62.13 & 76.74 & 75.29 & 54.61 & 74.26 & 77.68 \\
        SimSiam & 73.15 & 72.70 & 75.63 & 62.52 & 69.31 & 76.38 & 55.16 & 62.91 & 71.31 & 49.30 & 69.47 & 75.92 & 58.35 & 72.89 & 75.31 & 58.25 & 68.61 & 77.41 \\
        TS-TCC & 70.73 & 75.88 & 78.91 & 53.54 & 66.98 & 77.87 & 48.04 & 61.79 & 71.18 & 43.34 & 69.48 & 78.23 & 57.07 & 73.62 & 78.72 & 55.26 & 68.48 & 76.79 \\
        CLOCS & 68.94 & 73.36 & 76.31 & 57.94 & 72.55 & 76.24 & 51.97 & 57.96 & 72.65 & 47.19 & 71.88 & 76.31 & 59.59 & 77.78 & 77.49 & 54.38 & 71.93 & 76.13 \\
        ASTCL & 72.51 & 77.31 & 81.02 & 61.86 & 68.77 & 76.51 & 44.14 & 60.93 & 66.99 & 52.38 & 71.98 & 76.05 & 57.90 & 77.01 & 79.51 & 56.40 & 70.87 & 75.79 \\
        CRT & 69.68 & 78.24 & 77.24 & 61.98 & 70.82 & 78.67 & 46.41 & 59.49 & 68.73 & 47.44 & 73.52 & 74.41 & 58.01 & 76.43 & 82.03 & 56.21 & 73.70 & 78.80 \\
        ST-MEM & 61.12 & 66.87 & 71.36 & 54.12 & 57.86 & 63.59 & 55.71 & 59.99 & 66.07 & 51.12 & 65.44 & 74.85 & 56.69 & 63.32 & 70.39 & 59.77 & 66.87 & 71.36 \\
        MERL & 82.39 & 86.27 & 88.67 & 64.90 & 80.56 & 84.72 & 58.26 & 72.43 & 79.65 & 53.33 & 82.88 & 88.34 & 70.33 & 85.32 & 90.57 & 66.60 & 82.74& 87.95 \\
        MELP   & \textbf{85.82} & 87.61 & 87.87 & \textbf{79.22} & \underline{84.40} & \underline{87.46} & 63.41 & 76.71 & 83.30 & \underline{88.83} & \underline{94.65} & \underline{96.91} & \underline{88.54} & \underline{91.75} & 94.32 & 78.25 & 84.83 & 90.17 \\
        D-BETA & 83.15 & \underline{88.36} & \underline{90.11} & 77.74 & 82.92 & 85.15 & \underline{70.10} & \underline{78.91} & \underline{83.98} & 86.61 & 92.83 & 96.71 & 85.46 & 91.35 & \underline{94.92} & \underline{80.04} & \underline{87.36} & \underline{90.71} \\
        \textbf{FOCAL} & \underline{85.50} & \textbf{90.37} & \textbf{91.45} & \underline{78.24} & \textbf{86.36} & \textbf{90.39} & \textbf{71.83} & \textbf{79.35} & \textbf{85.36} & \textbf{89.52} & \textbf{95.01} & \textbf{98.01} & \textbf{89.32} & \textbf{93.33} & \textbf{96.52} & \textbf{81.92} & \textbf{88.03} & \textbf{92.27} \\
        \bottomrule
    \end{tabular}
    }
\end{table*}

\section{Experiments}
\subsection{Datasets}
We pre-train the FOCAL using the MIMIC-ECG \citep{gowmimic} dataset and test it on the PTB-XL \citep{wagner2020ptb}, CPSC2018 \citep{liu2018open}, and CSN \citep{zheng2022large} datasets, following the benchmark proposed by \citep{liu2024zero}. All the ECGs in the datasets are 12-lead recordings. The MIMIC-ECG dataset contains nearly 800,000 ECG-report pairs. To improve data quality, we excluded samples with an empty report or reports containing fewer than three words, removed reports without useful information, and discarded ECGs with unexpected situations. Details regarding the train:validation:test split and other dataset-specific information are provided in the Appendix.

\subsection{Implementation Details}
\textbf{Pre-training}: We use a randomly initialized ViT \citep{dosovitskiy2020image} with 120 non-overlapping patches for ECG encoding, and BioClinicalBERT \citep{alsentzer2019publicly} for text. Optimization is performed using AdamW (learning rate \(2 \times 10^{-5}\), weight decay \(1 \times 10^{-4}\)) with a batch size of 100, a cosine annealing scheduler, and a 10\% warmup. The temperature parameter \(t\) is initialized to \(\log 10\). Coarse FOCAL is pre-trained for 10 epochs on original reports, followed by 3 epochs for fine FOCAL on fine-grained reports. Candidate waveform features are queried via Qwen3-8B \citep{yang2025qwen3} accelerated by vLLM \citep{kwon2023efficient}, and filtered using a 0.95 confidence threshold.

\noindent \textbf{Downstream Tasks}: For zero-shot classification, the entire model is frozen, and probabilities are computed via the similarity between ECG embeddings and category text prompts. For linear probing, the ECG encoder remains frozen while a newly initialized linear classifier is trained using 1\%, 10\%, and 100\% data fractions. All task performances are measured by macro AUC.

\subsection{Zero-Shot Ability}
The zero-shot classification results across six datasets are presented in Table \ref{table:zero shot}. Our proposed FOCAL framework consistently outperforms all state-of-the-art baselines, achieving the highest Macro AUC on every evaluated dataset and subset. Notably, FOCAL achieves an average Macro AUC of 81.1\%, surpassing the strongest baseline, MELP \citep{wang2025token}, by a significant margin of 2.1\%.

A detailed examination reveals that FOCAL exhibits particularly strong performance on tasks requiring the detection of highly localized and complex waveform features, such as PTBXL-Rhythm and PTBXL-Form. This improvement directly validates our Optimal Transport-driven architecture, which accurately grounds specific pathological tags to precise ECG segments.

\subsection{Linear Evaluation}
We aim to evaluate the learned model transferability to downstream supervised tasks. We froze the ECG encoder and fine-tuned a randomly initialized linear classification head on the training data with binary cross-entropy loss. We compared a series of contrastive and generative self-supervised learning methods. Results in Table \ref{tab:linear_probe} show that FOCAL still achieves the best performances across all methods \citep{grill2020bootstrap,zbontar2021barlow,chen2021exploring,kiyasseh2021clocs,zhang2023self} in most scenarios. 

\section{Analysis}
In this section, we conduct a series of experiments to provide an in-depth analysis of FOCAL. All the reported metrics reflect the average zero-shot AUC across the six datasets described above.

\subsection{Ablation Study}
We conduct a comprehensive ablation study to validate FOCAL's three core components, with results in Table \ref{table:ablation}.

\textbf{Alignment Mechanisms.} Removing fine-grained alignment entirely causes a severe performance drop. Substituting Semi-UOT with standard Cross-Attention also degrades performance, validating that OT enforces sparse, precise mapping over attention's global smoothing. Furthermore, reverting to standard balanced OT (strict marginals) introduces severe noise by forcing uniform alignment mass onto uninformative background signals, highlighting the necessity of our semi-unbalanced relaxation for temporally sparse clinical data.

\textbf{Loss Objective.} Removing the semantic similarity matrix to revert to hard negative contrastive learning degrades performance. This confirms that strict binary pairings erroneously push apart semantically identical tags across different reports, a conflict effectively resolved by our semantic soft targets.

\textbf{Coarse-to-Fine Pipeline.} The Coarse FOCAL model trained solely on original reports underperforms, confirming the need for LLM waveform enrichment. Training the coarse model for 3 additional epochs without enrichment yielded no gains, indicating prior convergence and ruling out benefits from extended training alone. Additionally, training FOCAL from scratch on augmented reports achieves comparable performance but incurs significantly higher computational costs, validating the efficiency of our curriculum.

\begin{figure}[t]
    \centering
    \includegraphics[width=0.85\linewidth]{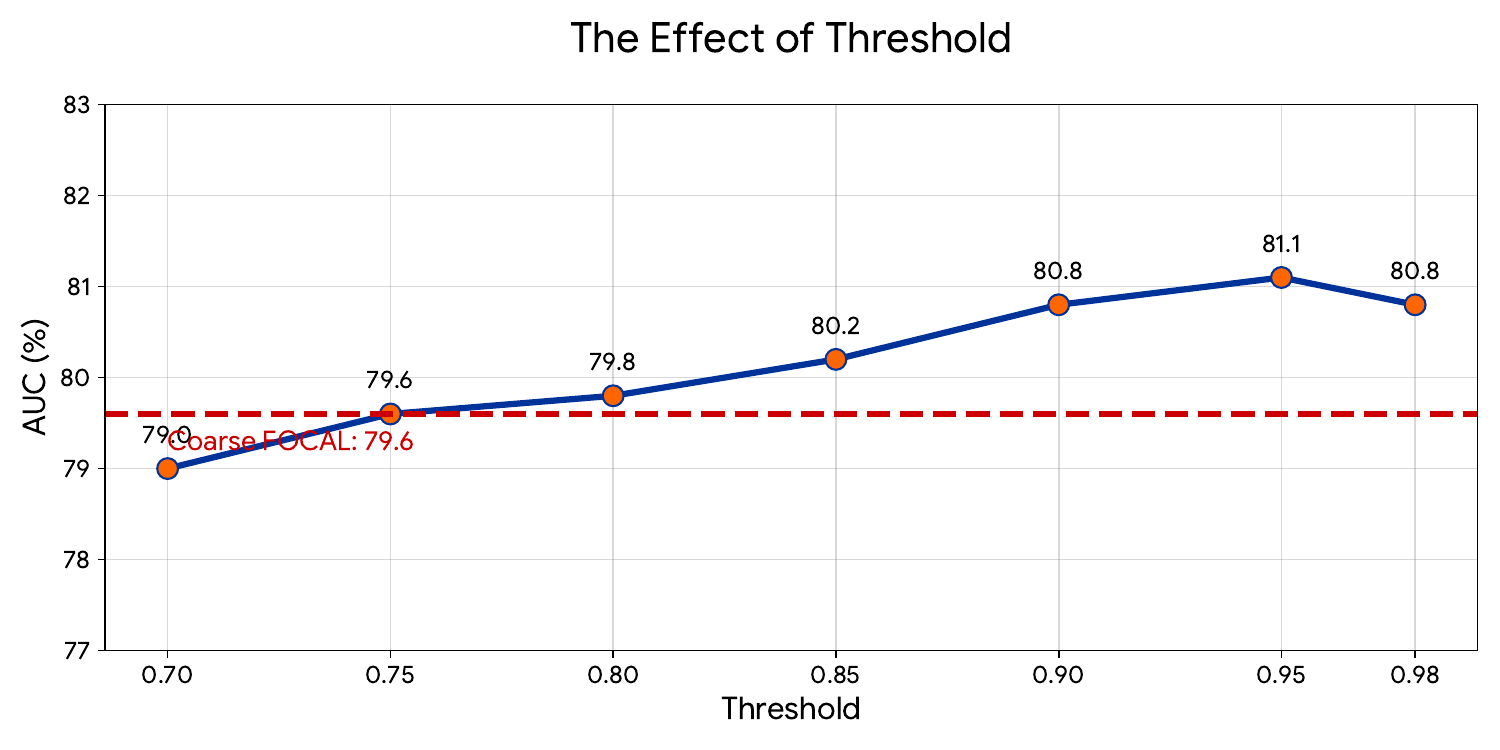}
    \caption{Effect of the filtering threshold}
    \label{fig:hyperparameter-effect}
\end{figure}

Finally, Figure \ref{fig:hyperparameter-effect} demonstrates FOCAL's robust performance across different LLM feature filtering thresholds. We retain a high default threshold of 0.95 to maximize precision and strictly filter out hallucinations.

\begin{table}[t]
    \centering
    \caption{Results of ablation study.}
    \label{table:ablation}
    \small
    \resizebox{0.9\linewidth}{!}{
    \begin{tabular}{lc}
        \toprule
        \textbf{Model Setting} & \textbf{AUC} \\ 
        \midrule
        \textbf{FOCAL (default)} & \textbf{81.1} \\
        \midrule
        \multicolumn{2}{l}{\textit{Alignment Mechanism Ablations}} \\
        w/ Cross-Attention (instead of OT) & 78.3 \\
        w/ Standard Balanced OT (Strict marginals) & 79.2 \\
        w/o Fine-Grained Alignment (Global only) & 76.0 \\
        \midrule
        \multicolumn{2}{l}{\textit{Loss Objective Ablations}} \\
        w/o Semantic Similarity Matrix & 77.6 \\
        \midrule
        \multicolumn{2}{l}{\textit{Training Pipeline Ablations}} \\
        w/o LLM Waveform Enrichment (coarse FOCAL) & 79.6 \\
        coarse FOCAL + 3 epochs (original reports) & 79.5 \\
        FOCAL (trained from scratch) & 81.1 \\
        \bottomrule
    \end{tabular}
    }
\end{table}

\begin{figure}[t]
    \centering
    \includegraphics[width=0.85\linewidth]{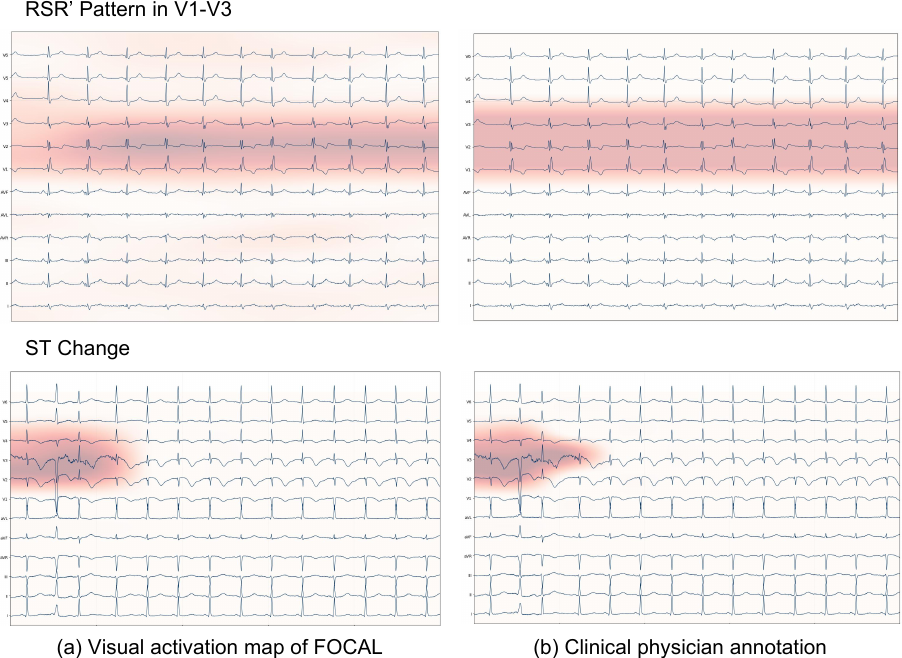}
    \caption{Visual activation maps generated by FOCAL.}
    \label{fig:visual_clinical}
\end{figure}

\begin{table}[t]
    \centering
    \caption{AUC of Generated Fine-Grained Reports (with/without verification).}
    \label{tab:ecg_auc}
    \small
    \begin{tabular}{lcc}
        \toprule
        \textbf{Waveform Feature} & \textbf{w/o Veri.} & \textbf{w Veri.} \\
        \midrule
        Non-specific ST elevation  & 71.42 & 83.42 \\
        Long QT-interval           & 84.25 & 93.25 \\
        Abnormal QRS               & 76.68 & 91.68 \\
        Prolonged PR interval      & 75.13 & 86.13 \\
        Inverted T-waves           & 74.79 & 88.79 \\
        \bottomrule
    \end{tabular}
\end{table}

\begin{table*}[t]
  \centering
  \small
  \caption{Impact of LLM-enriched reports on other baseline model (MERL).}
  \label{table:merl_enhancement}
  \resizebox{0.9\textwidth}{!}{
  \begin{tabular}{lccccccc}
    \toprule
    \textbf{Model Setting} & \textbf{PTBXL-Super} & \textbf{PTBXL-Sub} & \textbf{PTBXL-Form} & \textbf{PTBXL-Rhythm} & \textbf{CPSC2018} & \textbf{CSN} & \textbf{Average} \\
    \midrule
    MERL (Original Reports) & 74.2 & 75.7 & 65.9 & 78.5 & 82.8 & 74.4 & 75.3 \\
    MERL (w/ Enriched Reports) & 76.7 & 78.2 & 66.8 & 81.0 & 85.3 & 76.9 & 77.5 \\ 
    \midrule
    \textbf{Gain ($\Delta$)} & \textbf{+2.5} & \textbf{+2.5} & \textbf{+0.9} & \textbf{+2.5} & \textbf{+2.5} & \textbf{+2.5} & \textbf{+2.2} \\
    \bottomrule
  \end{tabular}
  }
\end{table*}

\begin{table*}[t]
\centering
\caption{Results under data distribution shift: `Source Domain' denotes the dataset used for linear probing with the frozen pre-trained ECG encoder. `Target Domain' refers to the corresponding test set. We include only those target domain samples that match categories from the source domain.}
    \scalebox{0.72}{    
    \begin{tabular}{l|c|c|cccccc}
        \toprule[1.2pt]
        \textbf{Source Domain} & \textbf{Zero-shot} & \textbf{Training Data} & \multicolumn{2}{c}{PTBXL-Super} & \multicolumn{2}{c}{CPSC2018} & \multicolumn{2}{c}{CSN} \\
        \textbf{Target Domain} & & \textbf{Ratio} & CPSC2018 & CSN & PTBXL-Super & CSN & PTBXL-Super & CPSC2018\\
        \midrule[1.2pt]
        Random Init & \xmark & 100\% & 68.62 & 75.31 & 55.74 & 68.92 & 56.57 & 61.16 \\
        SimCLR      & \xmark & 100\% & 69.62 & 73.05 & 56.65 & 66.36 & 59.74 & 62.11 \\
        BYOL        & \xmark & 100\% & 70.27 & 74.01 & 57.32 & 67.56 & 60.39 & 63.24\\
        BarlowTwins & \xmark & 100\% & 68.98 & 72.85 & 55.97 & 65.89 & 58.76 & 61.35 \\
        MoCo-v3     & \xmark & 100\% & 69.41 & 73.29 & 56.54 & 66.12 & 59.82 & 62.07\\
        SimSiam     & \xmark & 100\% & 70.06 & 73.92 & 57.21 & 67.48 & 60.23 & 63.09\\
        TS-TCC      & \xmark & 100\% & 71.32 & 75.16 & 58.47 & 68.34 & 61.55 & 64.48\\
        CLOCS       & \xmark & 100\% & 68.79 & 72.64 & 55.86 & 65.73 & 58.69 & 61.27\\
        ASTCL       & \xmark & 100\% & 69.23 & 73.18 & 56.61 & 66.27 & 59.74 & 62.12\\
        CRT         & \xmark & 100\% & 70.15 & 74.08 & 57.39 & 67.62 & 60.48 & 63.33\\
        ST-MEM      & \xmark & 100\% & 76.12 & \textbf{84.50} & 62.27 & 75.19 & 73.05 & 64.66 \\
        MERL & \cmark & 0\% & \textbf{88.21} & 78.01 & 76.77 & 76.56 & 74.15 & \underline{82.86}\\
        MELP         & \cmark & 0\% & \underline{87.75} & 74.11 & \underline{77.89} & 80.32 & 74.67 & 82.72 \\
        D-BETA       & \cmark & 0\% & 72.09 & 79.11 & 77.12 & \underline{82.91} & \underline{76.24} & 80.10 \\
        \midrule
        FOCAL & \cmark & 0\% & 85.33 & \underline{82.98} & \textbf{79.25} & \textbf{83.32} & \textbf{78.01} & \textbf{83.30} \\
        \bottomrule
    \end{tabular}
    }
    \label{tab: shift cls}
\end{table*}

\subsection{Visual Activation Map}
To further demonstrate the precise alignment between report tags and specific ECG segments, we present visual activation maps generated by FOCAL. We visualize the alignment weights for two highly localized pathological prompts in Figure~\ref{fig:visual_clinical}. First, using the text prompt `RSR' Pattern in V1-V3', which is a key morphological feature for diagnosing right bundle branch block, FOCAL accurately localizes the corresponding waveform anomalies within the exact leads. Second, for the prompt `ST segment changes', the model clearly highlights the specific temporal patches exhibiting abnormal elevations or depressions. 

\subsection{Validation and Transferability of Enriched Reports}
To evaluate our waveform enrichment pipeline's accuracy, we randomly sampled 200 MIMIC-ECG records originally lacking waveform descriptions. Three medical students independently annotated five key waveform features per ECG, establishing ground-truth labels via majority vote. To account for varying textual granularities, we calculated the AUC using the semantic similarity between the generated texts and target features. As Table~\ref{tab:ecg_auc} shows, unverified LLM-generated features yield low AUCs, confirming the risks of hallucinations and non-bijective clinical mappings. Conversely, verification by the Coarse FOCAL model substantially improves the AUC, demonstrating the pipeline's reliability in filtering hallucinations to generate high-fidelity reports.

Beyond internal accuracy, our augmentation serves as a universally beneficial data engine. To prove model-agnostic transferability, we pre-trained the state-of-the-art MERL baseline \citep{liu2024zero} using our enriched reports. Table \ref{table:merl_enhancement} shows this verified corpus yields consistent zero-shot improvements across all six downstream datasets, achieving a +2.2\% average gain over original texts. This plug-and-play capability confirms that recovering missing waveform semantics elevates overall pre-training corpus quality, extending its value beyond FOCAL to benefit the broader representation learning community.


\begin{table}[t]
  \centering
  \small
  \caption{Results of varying the LLM and Text Encoder components in our FOCAL framework.}
  \label{table:different_components}
  \resizebox{1.0\linewidth}{!}{
  \begin{tabular}{lcc}
    \toprule
    \textbf{Component Variant} & \textbf{Coarse FOCAL} & \textbf{FOCAL} $\uparrow$ \\ 
    \midrule
    \multicolumn{3}{c}{\textbf{Different Large Language Models}} \\
    Qwen3-8B \citep{yang2025qwen3} & 79.6 & 81.1 \\
    Qwen3-32B \citep{yang2025qwen3} & 79.6 & 81.2 \\
    Llama-3.1-8B-Instruct \citep{grattafiori2024llama} & 79.6 & 80.5 \\
    Gemma-3-12B-It \citep{team2024gemma} & 79.6 & 80.7 \\ 
    Meditron-7B \citep{chen2023meditron} & 79.6 & 81.0 \\ 
    medgemma-27b-text-it \citep{sellergren2025medgemma} & 79.6 & 81.4 \\
    \midrule
    \multicolumn{3}{c}{\textbf{Different Text Encoders}} \\
    BioClinicalBERT \citep{alsentzer2019publicly} & 79.6 & 81.1 \\
    PubMedBERT \citep{gu2021domain} & 79.7 & 80.9 \\
    Med-CPT \citep{jin2023medcpt} & 78.4 & 80.6 \\
    BioBERT \citep{lee2020biobert} & 78.6 & 80.5 \\
    \bottomrule
  \end{tabular}%
  }
\end{table}

\subsection{Robustness to Data Distribution Shifts}
To evaluate cross-domain generalizability, we tested models on unseen target domains following the experimental setting of MERL \citep{liu2024zero} (Table \ref{tab: shift cls}). Unlike traditional uni-modal SSL methods that require 100\% of source domain labels for linear probing, FOCAL and other multimodal baselines operate in a strict zero-shot setting. Despite this constraint, FOCAL achieves the best performance in four out of six cross-domain tasks and remains highly competitive in the others.

\subsection{Robustness Across Different Components}
We conducted experiments to evaluate the performance of our framework using various LLMs and text encoders. The results, presented in Table \ref{table:different_components}, demonstrate that FOCAL is highly robust across different configurations. Specifically, larger LLMs \citep{yang2025qwen3,team2024gemma,chen2023meditron,sellergren2025medgemma} yield only marginal gains, and domain-specific medical LLMs perform comparably to general-purpose ones. This suggests that general LLMs already possess sufficient foundational medical knowledge for waveform recovery. More importantly, our coarse FOCAL verification acts as a rigorous quality filter, standardizing the enriched reports regardless of the LLM used. Furthermore, FOCAL consistently improves upon coarse baselines across all tested text encoders \citep{alsentzer2019publicly, gu2021domain, jin2023medcpt, lee2020biobert}, confirming the strong generalizability.

\section{Conclusion}
In this paper, we propose FOCAL, a novel framework for fine-grained ECG-text contrastive learning. By formulating cross-modal routing via Optimal Transport, FOCAL explicitly grounds pathological tags to localized ECG patches. To resolve false negative collisions inherent in tag-level alignment, we integrated a semantic similarity matrix via soft targets. Furthermore, our coarse-to-fine pipeline reliably recovers missing waveform semantics while filtering hallucinations. Extensive experiments demonstrate that FOCAL achieves state-of-the-art zero-shot performance, and our generated waveform-enriched reports provide a universally transferable data engine for the broader representation learning community.

\clearpage
\section*{Limitations}
One primary limitation of our proposed framework is the computational and time overhead associated with the Large Language Model (LLM) waveform enrichment pipeline. Querying an LLM to generate potential waveform candidates for a large-scale dataset like MIMIC-ECG, which contains approximately 800,000 clinical reports, requires substantial inference compute. However, it is important to emphasize that this text augmentation process is strictly off-line and represents a one-time computational investment. Once the generated features are verified by the Coarse FOCAL model, the resulting fine-grained reports form a static, high-fidelity dataset. This enriched corpus not only facilitates the efficient training of our final model without adding any online training overhead, but it will also be open-sourced as a plug-and-play data asset. By providing this resource to the broader ECG representation learning community, the initial computational cost is effectively amortized over long-term, community-wide usage.

\section*{Acknowledgments}
This work was partially supported by the Technical Innovation Key Project of Zhejiang Province (2025C01128), the National Natural Science Foundation of China (Grant Nos. 82272129, 82572380) and the Natural Science Foundation of Zhejiang Province (LHZSZ25H090002).

\bibliography{custom}

@inproceedings{li2026anyecg,
  title={anyECG-chat: A Generalist ECG-MLLM for Flexible ECG Input and Multi-Task Understanding},
  author={Li, Haitao and Li, Ziyu and Mao, Yiheng and Liu, Ziyi and Sun, Zhoujian and Huang, Zhengxing},
  booktitle={Proceedings of the AAAI Conference on Artificial Intelligence},
  volume={40},
  number={1},
  pages={597--605},
  year={2026}
}

@article{ding2025ai,
  title={AI modeling photoplethysmography to electrocardiography useful for predicting cardiovascular disease},
  author={Ding, Zhengyao and Hu, Yujian and Li, Ziyu and Mao, Yiheng and Li, Haitao and Zhou, Dongchen and Chu, Xuesen and Yu, Long and Liu, Ziyi and Wu, Fei and others},
  journal={NPJ Digital Medicine},
  volume={9},
  number={1},
  pages={61},
  year={2025},
  publisher={Nature Publishing Group UK London}
}

@article{huang2023survey,
  title={A survey on hallucination in large language models: Principles, taxonomy, challenges, and open questions},
  author={Huang, Lei and Yu, Weijiang and Ma, Weitao and Zhong, Weihong and Feng, Zhangyin and Wang, Haotian and Chen, Qianglong and Peng, Weihua and Feng, Xiaocheng and Qin, Bing and others},
  journal={ACM Transactions on Information Systems},
  year={2023},
  publisher={ACM New York, NY}
}

@article{gunay2024accuracy,
  title={The accuracy of Gemini, GPT-4, and GPT-4o in ECG analysis: A comparison with cardiologists and emergency medicine specialists},
  author={G{\"u}nay, Serkan and {\"O}zt{\"u}rk, Ahmet and Yi{\u{g}}it, Yavuz},
  journal={The American journal of emergency medicine},
  volume={84},
  pages={68--73},
  year={2024},
  publisher={Elsevier}
}

@article{sahoo2020machine,
  title={Machine learning approach to detect cardiac arrhythmias in ECG signals: A survey},
  author={Sahoo, S and Dash, M and Behera, S and Sabut, S},
  journal={Irbm},
  volume={41},
  number={4},
  pages={185--194},
  year={2020},
  publisher={Elsevier}
}

@article{rath2021heart,
  title={Heart disease detection using deep learning methods from imbalanced ECG samples},
  author={Rath, Adyasha and Mishra, Debahuti and Panda, Ganapati and Satapathy, Suresh Chandra},
  journal={Biomedical Signal Processing and Control},
  volume={68},
  pages={102820},
  year={2021},
  publisher={Elsevier}
}

@article{ayano2022interpretable,
  title={Interpretable machine learning techniques in ECG-based heart disease classification: a systematic review},
  author={Ayano, Yehualashet Megersa and Schwenker, Friedhelm and Dufera, Bisrat Derebssa and Debelee, Taye Girma},
  journal={Diagnostics},
  volume={13},
  number={1},
  pages={111},
  year={2022},
  publisher={MDPI}
}

@article{jin2018screening,
  title={Screening for cardiovascular disease risk with ECG},
  author={Jin, Jill},
  journal={Jama},
  volume={319},
  number={22},
  pages={2346--2346},
  year={2018},
  publisher={American Medical Association}
}

@inproceedings{radford2021learning,
  title={Learning transferable visual models from natural language supervision},
  author={Radford, Alec and Kim, Jong Wook and Hallacy, Chris and Ramesh, Aditya and Goh, Gabriel and Agarwal, Sandhini and Sastry, Girish and Askell, Amanda and Mishkin, Pamela and Clark, Jack and others},
  booktitle={International conference on machine learning},
  pages={8748--8763},
  year={2021},
  organization={PMLR}
}

@inproceedings{li2024frozen,
  title={Frozen language model helps ecg zero-shot learning},
  author={Li, Jun and Liu, Che and Cheng, Sibo and Arcucci, Rossella and Hong, Shenda},
  booktitle={Medical Imaging with Deep Learning},
  pages={402--415},
  year={2024},
  organization={PMLR}
}

@article{liu2024zero,
  title={Zero-Shot ECG Classification with Multimodal Learning and Test-time Clinical Knowledge Enhancement},
  author={Liu, Che and Wan, Zhongwei and Ouyang, Cheng and Shah, Anand and Bai, Wenjia and Arcucci, Rossella},
  journal={arXiv preprint arXiv:2403.06659},
  year={2024}
}

@inproceedings{liu2024etp,
  title={Etp: Learning transferable ecg representations via ecg-text pre-training},
  author={Liu, Che and Wan, Zhongwei and Cheng, Sibo and Zhang, Mi and Arcucci, Rossella},
  booktitle={ICASSP 2024-2024 IEEE International Conference on Acoustics, Speech and Signal Processing (ICASSP)},
  pages={8230--8234},
  year={2024},
  organization={IEEE}
}

@article{yu2024ecg,
  title={ECG Semantic Integrator (ESI): A Foundation ECG Model Pretrained with LLM-Enhanced Cardiological Text},
  author={Yu, Han and Guo, Peikun and Sano, Akane},
  journal={arXiv preprint arXiv:2405.19366},
  year={2024}
}

@article{lalam2023ecg,
  title={Ecg representation learning with multi-modal ehr data},
  author={Lalam, Sravan Kumar and Kunderu, Hari Krishna and Ghosh, Shayan and Kumar, Harish and Awasthi, Samir and Prasad, Ashim and Lopez-Jimenez, Francisco and Attia, Zachi I and Asirvatham, Samuel and Friedman, Paul and others},
  journal={Transactions on Machine Learning Research},
  year={2023}
}

@inproceedings{chen2020simple,
  title={A simple framework for contrastive learning of visual representations},
  author={Chen, Ting and Kornblith, Simon and Norouzi, Mohammad and Hinton, Geoffrey},
  booktitle={International conference on machine learning},
  pages={1597--1607},
  year={2020},
  organization={PMLR}
}

@article{wang2023adversarial,
  title={Adversarial spatiotemporal contrastive learning for electrocardiogram signals},
  author={Wang, Ning and Feng, Panpan and Ge, Zhaoyang and Zhou, Yanjie and Zhou, Bing and Wang, Zongmin},
  journal={IEEE Transactions on Neural Networks and Learning Systems},
  year={2023},
  publisher={IEEE}
}

@article{eldele2021time,
  title={Time-series representation learning via temporal and contextual contrasting},
  author={Eldele, Emadeldeen and Ragab, Mohamed and Chen, Zhenghua and Wu, Min and Kwoh, Chee Keong and Li, Xiaoli and Guan, Cuntai},
  journal={arXiv preprint arXiv:2106.14112},
  year={2021}
}

@article{zhang2022maefe,
  title={Maefe: Masked autoencoders family of electrocardiogram for self-supervised pretraining and transfer learning},
  author={Zhang, Huaicheng and Liu, Wenhan and Shi, Jiguang and Chang, Sheng and Wang, Hao and He, Jin and Huang, Qijun},
  journal={IEEE Transactions on Instrumentation and Measurement},
  volume={72},
  pages={1--15},
  year={2022},
  publisher={IEEE}
}

@article{hu2023spatiotemporal,
  title={Spatiotemporal self-supervised representation learning from multi-lead ECG signals},
  author={Hu, Rui and Chen, Jie and Zhou, Li},
  journal={Biomedical Signal Processing and Control},
  volume={84},
  pages={104772},
  year={2023},
  publisher={Elsevier}
}

@article{zhang2022self,
  title={Self-Supervised Time Series Representation Learning via Cross Reconstruction Transformer},
  author={Zhang, Wenrui and Yang, Ling and Geng, Shijia and Hong, Shenda},
  journal={arXiv preprint arXiv:2205.09928},
  year={2022}
}

@inproceedings{zhai2023sigmoid,
  title={Sigmoid loss for language image pre-training},
  author={Zhai, Xiaohua and Mustafa, Basil and Kolesnikov, Alexander and Beyer, Lucas},
  booktitle={Proceedings of the IEEE/CVF International Conference on Computer Vision},
  pages={11975--11986},
  year={2023}
}

@article{gowmimic,
  title={MIMIC-IV-ECG-Diagnostic Electrocardiogram Matched Subset},
  author={Gow, Brian and Pollard, Tom and Nathanson, Larry A and Johnson, Alistair and Moody, Benjamin and Fernandes, Chrystinne and Greenbaum, Nathaniel and Berkowitz, Seth and Moukheiber, Dana and Eslami, Parastou and others}
}

@article{liu2018open,
  title={An open access database for evaluating the algorithms of electrocardiogram rhythm and morphology abnormality detection},
  author={Liu, Feifei and Liu, Chengyu and Zhao, Lina and Zhang, Xiangyu and Wu, Xiaoling and Xu, Xiaoyan and Liu, Yulin and Ma, Caiyun and Wei, Shoushui and He, Zhiqiang and others},
  journal={Journal of Medical Imaging and Health Informatics},
  volume={8},
  number={7},
  pages={1368--1373},
  year={2018},
  publisher={American Scientific Publishers}
}

@article{wagner2020ptb,
  title={PTB-XL, a large publicly available electrocardiography dataset},
  author={Wagner, Patrick and Strodthoff, Nils and Bousseljot, Ralf-Dieter and Kreiseler, Dieter and Lunze, Fatima I and Samek, Wojciech and Schaeffter, Tobias},
  journal={Scientific data},
  volume={7},
  number={1},
  pages={154},
  year={2020},
  publisher={Nature Publishing Group UK London}
}

@article{zheng2022large,
  title={A large scale 12-lead electrocardiogram database for arrhythmia study (version 1.0. 0)},
  author={Zheng, J and Guo, H and Chu, H},
  journal={PhysioNet 2022Available online: http://physionet. org/content/ecg-arrhythmia/1.0. 0/(accessed on 23 November 2022)},
  year={2022}
}

@article{alsentzer2019publicly,
  title={Publicly available clinical BERT embeddings},
  author={Alsentzer, Emily and Murphy, John R and Boag, Willie and Weng, Wei-Hung and Jin, Di and Naumann, Tristan and McDermott, Matthew},
  journal={arXiv preprint arXiv:1904.03323},
  year={2019}
}

@article{lavoie2024modeling,
  title={Modeling caption diversity in contrastive vision-language pretraining},
  author={Lavoie, Samuel and Kirichenko, Polina and Ibrahim, Mark and Assran, Mahmoud and Wildon, Andrew Gordon and Courville, Aaron and Ballas, Nicolas},
  journal={arXiv preprint arXiv:2405.00740},
  year={2024}
}

@article{jiang2023vision,
  title={Vision Lanauge Pre-training by Contrastive Learning with Cross-Modal Similarity Regulation},
  author={Jiang, Chaoya and Ye, Wei and Xu, Haiyang and Zhang, Shikun and Zhang, Jie and Huang, Fei and others},
  journal={arXiv preprint arXiv:2305.04474},
  year={2023}
}

@inproceedings{sun2023learning,
  title={Learning audio-visual source localization via false negative aware contrastive learning},
  author={Sun, Weixuan and Zhang, Jiayi and Wang, Jianyuan and Liu, Zheyuan and Zhong, Yiran and Feng, Tianpeng and Guo, Yandong and Zhang, Yanhao and Barnes, Nick},
  booktitle={Proceedings of the IEEE/CVF Conference on Computer Vision and Pattern Recognition},
  pages={6420--6429},
  year={2023}
}

@article{li2023integrating,
  title={Integrating language guidance into image-text matching for correcting false negatives},
  author={Li, Zheng and Guo, Caili and Feng, Zerun and Hwang, Jenq-Neng and Du, Zhongtian},
  journal={IEEE Transactions on Multimedia},
  year={2023},
  publisher={IEEE}
}

@article{wang2022medclip,
  title={Medclip: Contrastive learning from unpaired medical images and text},
  author={Wang, Zifeng and Wu, Zhenbang and Agarwal, Dinesh and Sun, Jimeng},
  journal={arXiv preprint arXiv:2210.10163},
  year={2022}
}

@inproceedings{kwon2023efficient,
  title={Efficient Memory Management for Large Language Model Serving with PagedAttention},
  author={Woosuk Kwon and Zhuohan Li and Siyuan Zhuang and Ying Sheng and Lianmin Zheng and Cody Hao Yu and Joseph E. Gonzalez and Hao Zhang and Ion Stoica},
  booktitle={Proceedings of the ACM SIGOPS 29th Symposium on Operating Systems Principles},
  year={2023}
}

@article{yang2025qwen3,
  title={Qwen3 technical report},
  author={Yang, An and Li, Anfeng and Yang, Baosong and Zhang, Beichen and Hui, Binyuan and Zheng, Bo and Yu, Bowen and Gao, Chang and Huang, Chengen and Lv, Chenxu and others},
  journal={arXiv preprint arXiv:2505.09388},
  year={2025}
}

@article{team2024gemma,
  title={Gemma: Open models based on gemini research and technology},
  author={Team, Gemma and Mesnard, Thomas and Hardin, Cassidy and Dadashi, Robert and Bhupatiraju, Surya and Pathak, Shreya and Sifre, Laurent and Rivi{\`e}re, Morgane and Kale, Mihir Sanjay and Love, Juliette and others},
  journal={arXiv preprint arXiv:2403.08295},
  year={2024}
}

@article{chen2023meditron,
  title={Meditron-70b: Scaling medical pretraining for large language models},
  author={Chen, Zeming and Cano, Alejandro Hern{\'a}ndez and Romanou, Angelika and Bonnet, Antoine and Matoba, Kyle and Salvi, Francesco and Pagliardini, Matteo and Fan, Simin and K{\"o}pf, Andreas and Mohtashami, Amirkeivan and others},
  journal={arXiv preprint arXiv:2311.16079},
  year={2023}
}

@article{sellergren2025medgemma,
  title={Medgemma technical report},
  author={Sellergren, Andrew and Kazemzadeh, Sahar and Jaroensri, Tiam and Kiraly, Atilla and Traverse, Madeleine and Kohlberger, Timo and Xu, Shawn and Jamil, Fayaz and Hughes, C{\'\i}an and Lau, Charles and others},
  journal={arXiv preprint arXiv:2507.05201},
  year={2025}
}

@article{gu2021domain,
  title={Domain-specific language model pretraining for biomedical natural language processing},
  author={Gu, Yu and Tinn, Robert and Cheng, Hao and Lucas, Michael and Usuyama, Naoto and Liu, Xiaodong and Naumann, Tristan and Gao, Jianfeng and Poon, Hoifung},
  journal={ACM Transactions on Computing for Healthcare (HEALTH)},
  volume={3},
  number={1},
  pages={1--23},
  year={2021},
  publisher={ACM New York, NY}
}

@article{jin2023medcpt,
  title={MedCPT: Contrastive Pre-trained Transformers with large-scale PubMed search logs for zero-shot biomedical information retrieval},
  author={Jin, Qiao and Kim, Won and Chen, Qingyu and Comeau, Donald C and Yeganova, Lana and Wilbur, W John and Lu, Zhiyong},
  journal={Bioinformatics},
  volume={39},
  number={11},
  pages={btad651},
  year={2023},
  publisher={Oxford University Press}
}

@article{lee2020biobert,
  title={BioBERT: a pre-trained biomedical language representation model for biomedical text mining},
  author={Lee, Jinhyuk and Yoon, Wonjin and Kim, Sungdong and Kim, Donghyeon and Kim, Sunkyu and So, Chan Ho and Kang, Jaewoo},
  journal={Bioinformatics},
  volume={36},
  number={4},
  pages={1234--1240},
  year={2020},
  publisher={Oxford University Press}
}

@article{dosovitskiy2020image,
  title={An image is worth 16x16 words: Transformers for image recognition at scale},
  author={Dosovitskiy, Alexey and Beyer, Lucas and Kolesnikov, Alexander and Weissenborn, Dirk and Zhai, Xiaohua and Unterthiner, Thomas and Dehghani, Mostafa and Minderer, Matthias and Heigold, Georg and Gelly, Sylvain and others},
  journal={arXiv preprint arXiv:2010.11929},
  year={2020}
}

@article{grill2020bootstrap,
  title={Bootstrap your own latent-a new approach to self-supervised learning},
  author={Grill, Jean-Bastien and Strub, Florian and Altch{\'e}, Florent and Tallec, Corentin and Richemond, Pierre and Buchatskaya, Elena and Doersch, Carl and Avila Pires, Bernardo and Guo, Zhaohan and Gheshlaghi Azar, Mohammad and others},
  journal={Advances in neural information processing systems},
  volume={33},
  pages={21271--21284},
  year={2020}
}

@inproceedings{zbontar2021barlow,
  title={Barlow twins: Self-supervised learning via redundancy reduction},
  author={Zbontar, Jure and Jing, Li and Misra, Ishan and LeCun, Yann and Deny, St{\'e}phane},
  booktitle={International conference on machine learning},
  pages={12310--12320},
  year={2021},
  organization={PMLR}
}

@inproceedings{chen2021empirical,
  title={An empirical study of training self-supervised vision transformers},
  author={Chen, Xinlei and Xie, Saining and He, Kaiming},
  booktitle={Proceedings of the IEEE/CVF international conference on computer vision},
  pages={9640--9649},
  year={2021}
}

@inproceedings{chen2021exploring,
  title={Exploring simple siamese representation learning},
  author={Chen, Xinlei and He, Kaiming},
  booktitle={Proceedings of the IEEE/CVF conference on computer vision and pattern recognition},
  pages={15750--15758},
  year={2021}
}

@inproceedings{kiyasseh2021clocs,
  title={Clocs: Contrastive learning of cardiac signals across space, time, and patients},
  author={Kiyasseh, Dani and Zhu, Tingting and Clifton, David A},
  booktitle={International Conference on Machine Learning},
  pages={5606--5615},
  year={2021},
  organization={PMLR}
}

@article{zhang2023self,
  title={Self-supervised time series representation learning via cross reconstruction transformer},
  author={Zhang, Wenrui and Yang, Ling and Geng, Shijia and Hong, Shenda},
  journal={IEEE Transactions on Neural Networks and Learning Systems},
  year={2023},
  publisher={IEEE}
}

@article{na2024guiding,
  title={Guiding Masked Representation Learning to Capture Spatio-Temporal Relationship of Electrocardiogram},
  author={Na, Yeongyeon and Park, Minje and Tae, Yunwon and Joo, Sunghoon},
  journal={arXiv preprint arXiv:2402.09450},
  year={2024}
}

@article{liang2025medfilip,
  title={Medfilip: Medical fine-grained language-image pre-training},
  author={Liang, Xinjie and Li, Xiangyu and Li, Fanding and Jiang, Jie and Dong, Qing and Wang, Wei and Wang, Kuanquan and Dong, Suyu and Luo, Gongning and Li, Shuo},
  journal={IEEE Journal of Biomedical and Health Informatics},
  year={2025},
  publisher={IEEE}
}

@article{shui2025large,
  title={Large-scale and fine-grained vision-language pre-training for enhanced ct image understanding},
  author={Shui, Zhongyi and Zhang, Jianpeng and Cao, Weiwei and Wang, Sinuo and Guo, Ruizhe and Lu, Le and Yang, Lin and Ye, Xianghua and Liang, Tingbo and Zhang, Qi and others},
  journal={arXiv preprint arXiv:2501.14548},
  year={2025}
}

@article{kim2025falcon,
  title={FALCON: False-Negative Aware Learning of Contrastive Negatives in Vision-Language Pretraining},
  author={Kim, Myunsoo and Shim, Seong-Woong and Lee, Byung-Jun},
  journal={arXiv preprint arXiv:2505.11192},
  year={2025}
}

@article{ni2025towards,
  title={Towards trustworthy retrieval augmented generation for large language models: A survey},
  author={Ni, Bo and Liu, Zheyuan and Wang, Leyao and Lei, Yongjia and Zhao, Yuying and Cheng, Xueqi and Zeng, Qingkai and Dong, Luna and Xia, Yinglong and Kenthapadi, Krishnaram and others},
  journal={arXiv preprint arXiv:2502.06872},
  year={2025}
}

@article{gao2023retrieval,
  title={Retrieval-augmented generation for large language models: A survey},
  author={Gao, Yunfan and Xiong, Yun and Gao, Xinyu and Jia, Kangxiang and Pan, Jinliu and Bi, Yuxi and Dai, Yixin and Sun, Jiawei and Wang, Haofen and Wang, Haofen},
  journal={arXiv preprint arXiv:2312.10997},
  volume={2},
  number={1},
  year={2023}
}

@inproceedings{wang2025token,
  title={From Token to Rhythm: A Multi-Scale Approach for ECG-Language Pretraining},
  author={Wang, Fuying and Xu, Jiacheng and Yu, Lequan},
  booktitle={International Conference on Machine Learning},
  pages={65059--65074},
  year={2025},
  organization={PMLR}
}

@inproceedings{hung2025boosting,
  title={Boosting Masked ECG-Text Auto-Encoders as Discriminative Learners},
  author={Hung, Manh Pham and Saeed, Aaqib and Ma, Dong},
  booktitle={Forty-second International Conference on Machine Learning},
  year={2025}
}

@article{grattafiori2024llama,
  title={The llama 3 herd of models},
  author={Grattafiori, Aaron and Dubey, Abhimanyu and Jauhri, Abhinav and Pandey, Abhinav and Kadian, Abhishek and Al-Dahle, Ahmad and Letman, Aiesha and Mathur, Akhil and Schelten, Alan and Vaughan, Alex and others},
  journal={arXiv preprint arXiv:2407.21783},
  year={2024}
}
\clearpage

\appendix

\section{Dataset Analysis}
\label{appendix:data}
We pre-train the FOCAL using the MIMIC-ECG dataset and test it on the PTB-XL, CPSC2018, and CSN datasets. All the ECGs in the datasets are 12-lead recordings. The PTB-XL dataset can be further divided into four subsets, and we follow the official train:validation:test split. For CPSC2018 and CSN, we split the dataset as 70\%:10\%:20\% for the train:validation:test split. The statistics of the datasets used are presented in Table \ref{table:dataset}.

\textbf{MIMIC-ECG} The MIMIC-ECG dataset contains nearly 800,000 ECG-report pairs from approximately 160,000 unique patients. These diagnostic ECGs utilize 12 leads and are 10 seconds in duration, with a sampling rate of 500 Hz. 

\textbf{PTB-XL} The PTB-XL ECG dataset is a large dataset of 21,799 clinical 12-lead ECGs from 18,869 patients of 10-second length. There are four subsets with multi-label classification tasks: Superclass (5 categories), Subclass (23 categories), Form (19 categories), and Rhythm (12 categories). Notably, these four subsets have different numbers of samples. 

\textbf{CPSC2018} This publicly accessible dataset comprises 6,877 standard 12-lead ECG records, each sampled at a rate of 500 Hz, with durations ranging from 6 to 60 seconds. The dataset is annotated with 9 distinct labels. 

\textbf{Chapman-Shaoxing-Ningbo (CSN)} This dataset contains huge 12-lead ECGs with a 500 Hz sampling rate, which features multiple common rhythms and additional cardiovascular conditions, all labeled by professional experts. 

\begin{table}[htb]
  \centering
  \small
  \caption{The statistics of used datasets.}
  \resizebox{1.0\columnwidth}{!}{%
    \begin{tabular}{lcccc}
      \toprule
      \textbf{Pretrain} & \textbf{\# ECGs} &  \textbf{\# Reports}  \\ 
      \midrule
      MIMIC-ECG & 773,268 & 773,268 \\
      \toprule
      \textbf{Evaluation} & \textbf{\# Train} & \textbf{\# Valid} &  \textbf{\# Test} &  \textbf{\# Classes} \\
      \midrule
      PTBXL-Super & 17,084 & 2,146 & 2,158 & 5\\
      PTBXL-Sub & 17,084 & 2,146 & 2,158 & 23\\
      PTBXL-Form & 7,197 & 901 & 880 & 19\\
      PTBXL-Rhythm & 16,832 & 2,100 & 2,098 & 12\\
      CPSC2018 & 4,950 & 551 & 1,376 & 9\\
      CSN & 16,546 & 1,860 & 4,620 & 38\\
      \bottomrule
    \end{tabular}%
  }
  \label{table:dataset}
\end{table}

\section{Pseudo Code}
\label{appedix:pseudo_code}
The pseudo-code of our FOCAL training pipeline is summarized in Algorithm \ref{alg:pseudo_code}.

\begin{algorithm}
\small
\caption{Coarse-to-Fine Training Pipeline of FOCAL}\label{alg:pseudo_code}
\begin{algorithmic}[1]
\STATE \textbf{Input:} Original dataset \( D = \{(x_{\text{ecg}_i}, x_{\text{txt}_i}) \mid i \in [0, n)\} \), LLM prompt, threshold \( \tau \)
\STATE \textbf{Output:} Optimized FOCAL model
\STATE \textit{\# Step 1: Train Coarse Model}
\STATE Train Coarse FOCAL model \( f_{\text{coarse}} \) using original dataset \( D \)
\STATE \textit{\# Step 2: LLM Waveform Enrichment \& Verification}
\STATE Initialize augmented dataset \( D_{\text{aug}} = \emptyset \)
\FOR{$i = 0 \text{ to } n-1$}
    \STATE \( \text{features} = \text{LLM}(x_{\text{txt}_i}, \text{prompt}) \) \quad \textit{\# Generate candidate waveform features}
    \STATE \( x_{\text{txt}_i}^{\text{aug}} = x_{\text{txt}_i} \)
    \FOR{each \( f_j \in \text{features} \)}
        \STATE \( \text{sim} = f_{\text{coarse}}(x_{\text{ecg}_i}, f_j) \) \quad \textit{\# Compute cross-modal similarity}
        \IF{\( \text{sim} > \tau \)}
            \STATE \( x_{\text{txt}_i}^{\text{aug}} = x_{\text{txt}_i}^{\text{aug}} \cup \{f_j\} \) \quad \textit{\# Retain high-confidence feature}
        \ENDIF
    \ENDFOR
    \STATE \( D_{\text{aug}} = D_{\text{aug}} \cup \{(x_{\text{ecg}_i}, x_{\text{txt}_i}^{\text{aug}})\} \)
\ENDFOR
\STATE \textit{\# Step 3: Final FOCAL Training}
\STATE Continue training \( f_{\text{coarse}} \) on augmented dataset \( D_{\text{aug}} \) to obtain final FOCAL
\end{algorithmic}
\end{algorithm}

\section{Semantic Similarity Matrix}
\begin{figure}[h]
    \centering
    \includegraphics[width=0.8\linewidth]{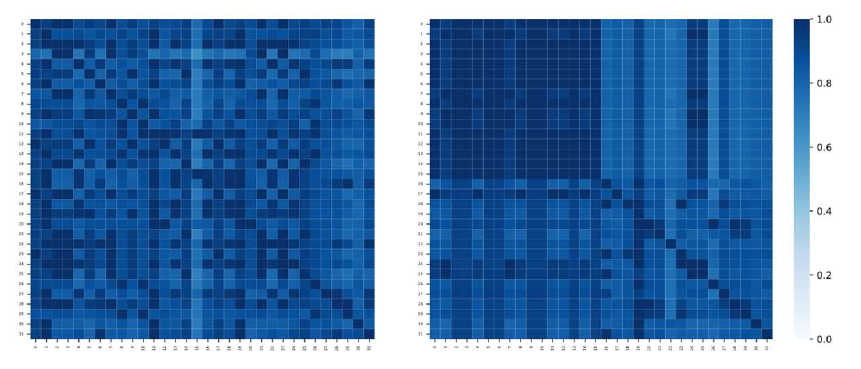} 
    \caption{\textbf{The Heatmap of Semantic Similarity Matrix.} 
    Left: a random batch; Right: with the first 16 as normal ECG and the last 16 as abnormal ECG.}
    \label{fig:heatmap}
\end{figure}

We visualize the semantic similarity matrix in Figure \ref{fig:heatmap}. The left side shows the semantic similarity matrix from a random batch. As illustrated, ECGs and tags from different records may share similarities to some extent. Ignoring these similarities would result in a diagonal matrix with ones on the diagonal and zeros elsewhere, which is obviously wrong. The right side displays a semantic similarity matrix where the first 16 entries are normal ECGs and the last 16 are abnormal ECGs. The matrix effectively captures the semantic similarities of the normal ECGs. 

\begin{table*}[t]
  \centering
  \small
  \caption{Relative Inference Time Comparison across baselines.}
  \label{tab:inference_time}
  \resizebox{0.95\linewidth}{!}{%
  \begin{tabular}{lccl}
    \toprule
    \textbf{Method} & \textbf{LLM at Inference} & \textbf{Relative Time} $\downarrow$ & \textbf{Notes} \\
    \midrule
    MERL \citep{liu2024zero} & \cmark (CKEPE) & 1.17$\times$ & LLM call required per sample at test time \\
    MELP \citep{wang2025token} & \xmark  & 1.08$\times$ & Overhead from multi-scale encoder \\
    D-BETA \citep{hung2025boosting} & \xmark  & 0.98$\times$ & MAE used only during training \\
    \textbf{FOCAL (Ours)} & \xmark  & 1.00$\times$ & Sinkhorn-Knopp adds negligible overhead \\
    \bottomrule
  \end{tabular}%
  }
\end{table*}

\begin{table*}[t]
  \centering
  \caption{Full $2 \times 2$ Ablation of Fine-Grained Architecture and Waveform Enrichment (Zero-Shot Avg. AUC \%).}
  \label{tab:2x2_ablation}
  \resizebox{\linewidth}{!}{%
  \begin{tabular}{lccc}
    \toprule
    \textbf{Method} & \textbf{Fine-Grained Architecture} & \textbf{Waveform Enrichment} & \textbf{Avg. AUC (\%)} \\
    \midrule
    Global only (w/o both) & \xmark & \xmark & 74.8 \\
    w/o Fine-Grained Architecture$^\dagger$ & \xmark & \cmark & 76.0 \\
    Coarse FOCAL$^\dagger$ & \cmark & \xmark & 79.6 \\
    \textbf{FOCAL}$^\dagger$ & \cmark & \cmark & \textbf{81.1} \\
    \bottomrule
  \end{tabular}%
  }
  
  \vspace{2mm} 
\end{table*}

\section{Stability of the Semantic Similarity Matrix}
\label{sec:stability_S}
The semantic similarity matrix $\mathbf{S}$ is computed online from the text embeddings throughout the training process. A potential concern with online updates is whether the instability of the text encoder in the early stages of training could negatively affect the contrastive learning objective. In our framework, the text encoder is initialized from BioClinicalBERT \citep{alsentzer2019publicly} weights, which are pre-trained on large-scale biomedical text. Consequently, these embeddings encode reliable semantic relationships among clinical terms from the very first training step. This robust initialization ensures that the online computation of $\mathbf{S}$ remains stable and consistently provides high-quality soft targets, preventing any training collapse or instability.

\section{Inference-Time Efficiency Analysis}
\label{sec:efficiency}
To evaluate the computational overhead of our framework, we compare the relative inference time of FOCAL against key baselines in Table \ref{tab:inference_time}. We distinguish between training-time and inference-time costs. The LLM-based waveform enrichment in FOCAL is a strictly offline, one-time preprocessing step with zero online overhead. At inference time, the only addition is the unbalanced Sinkhorn-Knopp algorithm. Because this consists entirely of highly parallelizable matrix operations, its computational cost is negligible relative to standard neural network forward passes. As shown in Table \ref{tab:inference_time}, FOCAL is highly efficient, whereas methods like MERL \citep{liu2024zero} require live LLM calls per sample via their CKEPE pipeline during inference.

\section{$2 \times 2$ Ablation of Fine-Grained Architecture and Enrichment}
\label{sec:2x2_ablation}
To further disentangle the performance gains contributed by our architectural design versus the data augmentation strategy, we provide a full $2 \times 2$ ablation study in Table \ref{tab:2x2_ablation}. The results clearly demonstrate that the fine-grained Semi-UOT architecture is the dominant contributor to our framework's success (yielding a $5.1\%$ absolute improvement over the global baseline). Meanwhile, the LLM waveform enrichment provides a consistent additive gain ($+1.5\%$) regardless of the underlying alignment configuration. This confirms that the improvements from the architecture and the enriched training text are orthogonal and complementary.

\section{Contribution of Alignment and Enrichment to Cross-Domain Transfer}
\label{sec:cross_domain_ablation}
To complement the evaluation and explicitly address how our core mechanisms affect generalization, we dissect the contributions of alignment and enrichment in the cross-domain transfer setting. Table \ref{tab:cross_domain_ablation} shows that the fine-grained OT-driven alignment is the primary driver of cross-domain transferability. Removing it causes a $5\% \sim 6\%$ drop across all tasks. Text enrichment provides an additional $1\% \sim 2\%$ gain by improving the quality of the pre-training corpus.

\begin{table*}[t]
  \centering
  \small
  \caption{Ablation of Alignment and Enrichment on Cross-Domain Transfer (Zero-Shot AUC). "PTB" denotes PTBXL-Super.}
  \label{tab:cross_domain_ablation}
  \resizebox{1.0\textwidth}{!}{%
  \begin{tabular}{lcc|cccccc}
    \toprule
    \textbf{Method} & \textbf{Alignment} & \textbf{Enrichment} & \textbf{PTB$\rightarrow$CPSC} & \textbf{PTB$\rightarrow$CSN} & \textbf{CPSC$\rightarrow$PTB} & \textbf{CPSC$\rightarrow$CSN} & \textbf{CSN$\rightarrow$PTB} & \textbf{CSN$\rightarrow$CPSC} \\
    \midrule
    w/o Alignment & \xmark & \cmark & 79.51 & 76.83 & 73.21 & 77.61 & 72.53 & 77.42 \\
    Coarse FOCAL & \cmark & \xmark & 83.52 & 81.24 & 77.82 & 81.91 & 76.44 & 81.83 \\
    \textbf{FOCAL} & \cmark & \cmark & \textbf{85.33} & \textbf{82.98} & \textbf{79.25} & \textbf{83.32} & \textbf{78.01} & \textbf{83.30} \\
    \bottomrule
  \end{tabular}%
  }
\end{table*}

\section{Hyperparameter Sensitivity Analysis}
\label{sec:sensitivity}
We evaluate the robustness of our Semi-UOT formulation by varying its two core hyperparameters: the entropic regularization weight $\epsilon$ and the KL divergence penalty $\tau$. As demonstrated in Table \ref{tab:sensitivity_epsilon} and Table \ref{tab:sensitivity_tau}, FOCAL's performance remains highly stable, fluctuating by less than $0.5\%$ across a wide practical operating range. Extreme boundary cases predictably degrade performance: as $\epsilon \to 0$, the formulation approaches computationally unstable exact OT, and as $\tau \to \infty$, it reverts to standard balanced OT, which erroneously forces uniform alignment mass onto uninformative ECG baselines.

\begin{table}[h]
  \centering
  \small
  \caption{Sensitivity to Entropic Regularization $\epsilon$ (with $\tau = 1.0$ fixed).}
  \label{tab:sensitivity_epsilon}
  \resizebox{\linewidth}{!}{%
  \begin{tabular}{lccccc}
    \toprule
    $\epsilon$ & 0.01 & 0.05 & \textbf{0.1 (default)} & 0.2 & 0.5 \\
    \midrule
    \textbf{Avg. AUC (\%)} & 79.8 & 80.6 & \textbf{81.1} & 80.9 & 80.3 \\
    \bottomrule
  \end{tabular}%
  }
\end{table}

\begin{table}[h]
  \centering
  \small
  \caption{Sensitivity to KL Divergence Penalty $\tau$ (with $\epsilon = 0.1$ fixed).}
  \label{tab:sensitivity_tau}
  \resizebox{\linewidth}{!}{%
  \begin{tabular}{lccccc}
    \toprule
    $\tau$ & 0.1 & 0.5 & \textbf{1.0 (default)} & 2.0 & 5.0 \\
    \midrule
    \textbf{Avg. AUC (\%)} & 80.2 & 80.7 & \textbf{81.1} & 80.8 & 80.1 \\
    \bottomrule
  \end{tabular}%
  }
\end{table}

\section{Notation Table}
\label{appendix:notation}
Table \ref{tab:notation} summarizes the key mathematical notations used in this paper.

\begin{table*}[h]
  \centering
  \caption{Summary of Key Notations in FOCAL}
  \label{tab:notation}
  \small
  \begin{tabular}{ll}
    \toprule
    \textbf{Symbol} & \textbf{Description} \\
    \midrule
    $\mathcal{D}$ & The ECG-report multimodal dataset \\
    $\mathbf{X}_i$ & The $i$-th 12-lead ECG signal \\
    $\mathcal{R}_i$ & The clinical report paired with $\mathbf{X}_i$ \\
    $L, T$ & Number of leads ($L=12$) and temporal length of the ECG signal \\
    $N_{time}$ & Number of non-overlapping patches per lead \\
    $N$ & Total number of ECG patches ($L \times N_{time}$) \\
    $\mathbf{x}_i^p$ & The $p$-th flattened ECG patch \\
    $d$ & Dimension of the latent projection space \\
    $\mathbf{e}_{lead}^p, \mathbf{e}_{time}^p$ & Learnable lead and temporal embeddings for the $p$-th patch \\
    $\mathbf{H}_i, \mathbf{h}_i^p$ & Fine-grained patch embedding matrix and the $p$-th patch embedding \\
    $\mathcal{T}_i$ & Set of independent tags extracted from report $\mathcal{R}_i$ \\
    $m_i$ & Number of clinical tags in report $\mathcal{R}_i$ \\
    $\mathbf{Z}_i, \mathbf{z}_i^j$ & Tag-level representation matrix and the $j$-th tag embedding \\
    $\mathbf{C}, C_{j,p}$ & Cost matrix and the cosine distance between the $j$-th tag and $p$-th patch \\
    $\mathbf{T}, \hat{T}_{j,p}$ & Optimal transport plan and its normalized form \\
    $\mathbf{a}, \mathbf{b}$ & Prior marginal distributions for tags and ECG patches \\
    $\epsilon, \tau$ & Hyperparameters for entropic regularization and KL divergence penalty \\
    $\mathbf{e}_i^j$ & Tag-specific ECG representation routed via Semi-UOT \\
    $B, K$ & Batch size and the total number of tags in a batch \\
    $\mathbf{S}$ & Semantic similarity matrix computed from text tag embeddings \\
    $t$ & Learnable temperature parameter for contrastive learning \\
    \bottomrule
  \end{tabular}
\end{table*}

\section{Running Cases for Generating Fine-Grained Reports}
\begin{figure*}[t]
    \centering
    \includegraphics[width=0.8\linewidth]{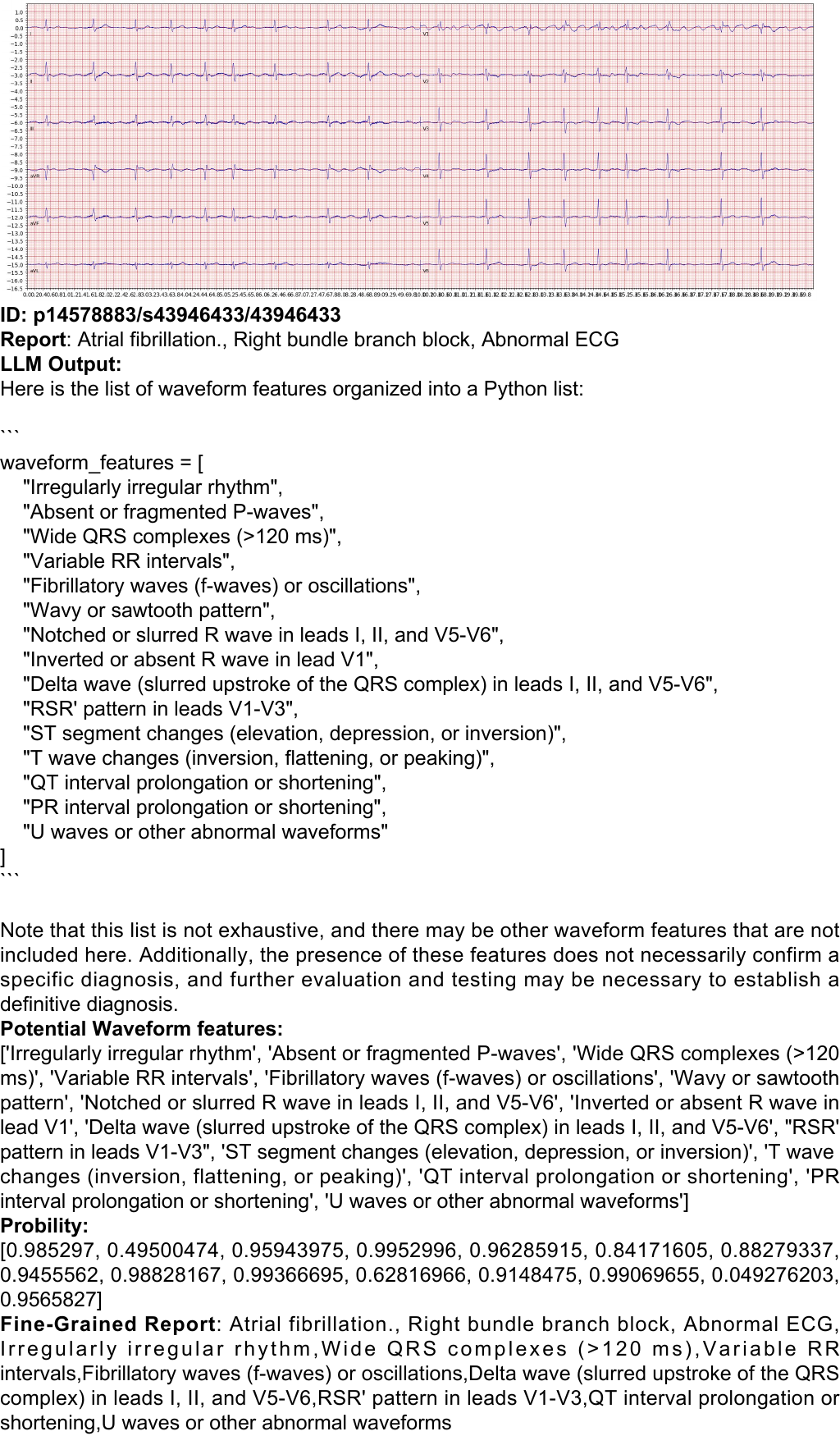}
    \caption{Running Case1 with Atrial fibrillation and Right bundle branch block.}
    \label{fig:case1}
\end{figure*}

\begin{figure*}[t]
    \centering
    \includegraphics[width=0.75\linewidth]{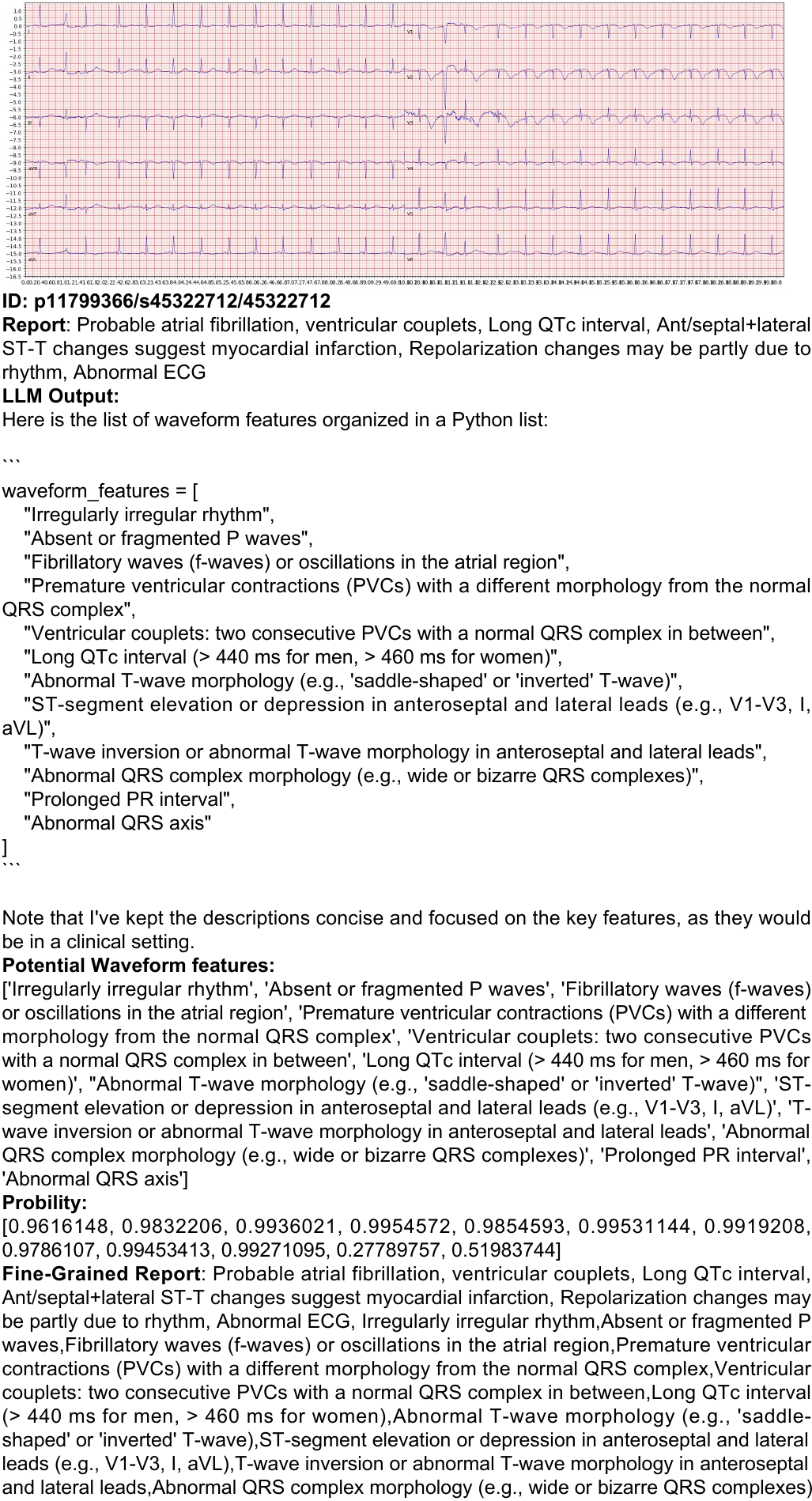}
    \caption{Running Case2 with Myocardial Infarction.}
    \label{fig:case2}
\end{figure*}

\begin{figure*}[t]
    \centering
    \includegraphics[width=0.9\linewidth]{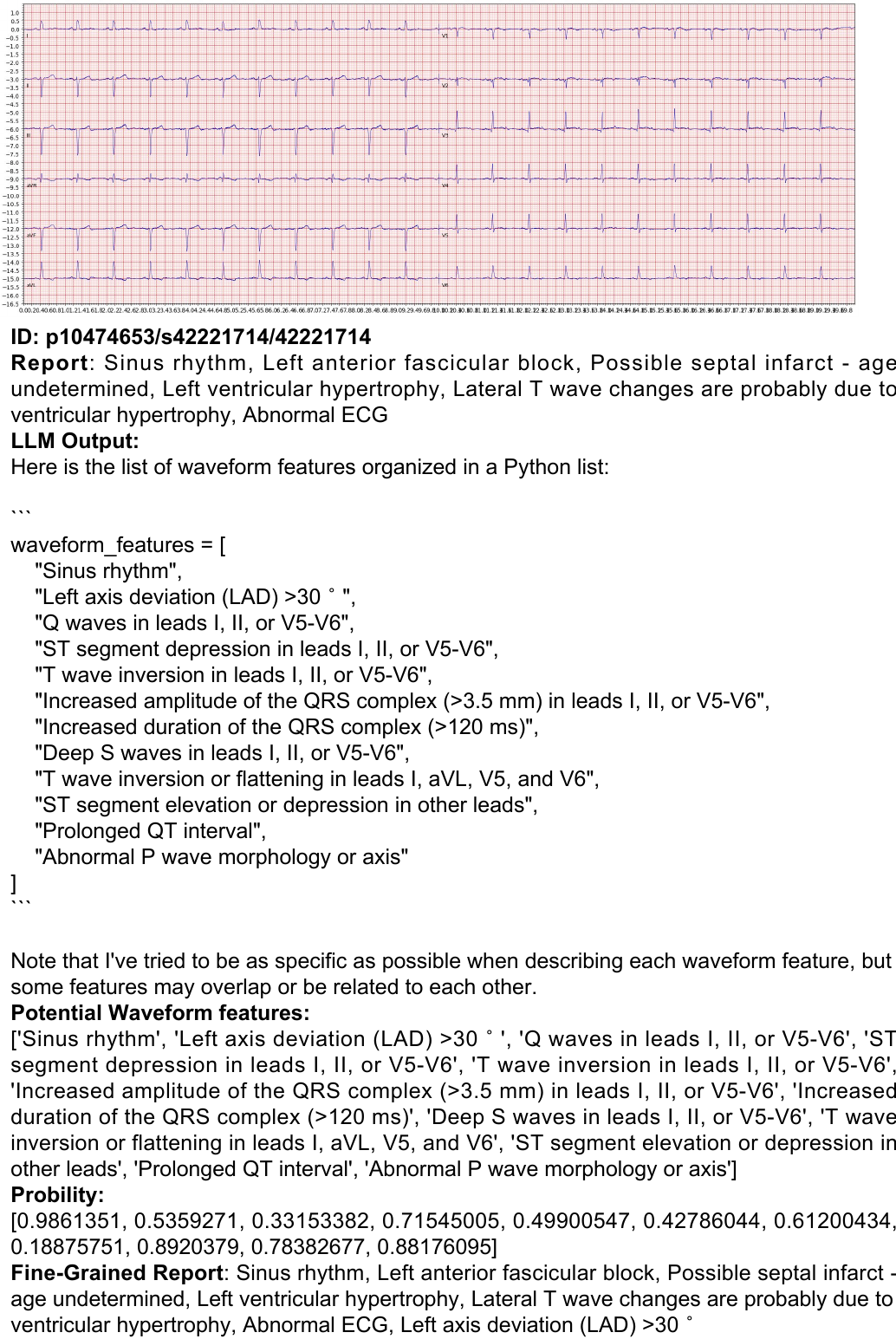}
    \caption{Running Case3 with Hypertrophy.}
    \label{fig:case3}
\end{figure*}

We present three case studies illustrating how the fine-grained reports with waveform features are generated step by step, as shown in Figure \ref{fig:case1}, \ref{fig:case2}, \ref{fig:case3}. The LLM is prompted with two explicit chain-of-thought instructions: \textit{`What waveform features are most likely to be present in electrocardiograms with these symptoms?'} followed by \textit{`Organize these waveform features into a Python list, with each item representing a distinct waveform feature.'}

\section{ECG-Text Retrieval}
\begin{figure*}[t]
    \centering
    \includegraphics[width=0.8\linewidth]{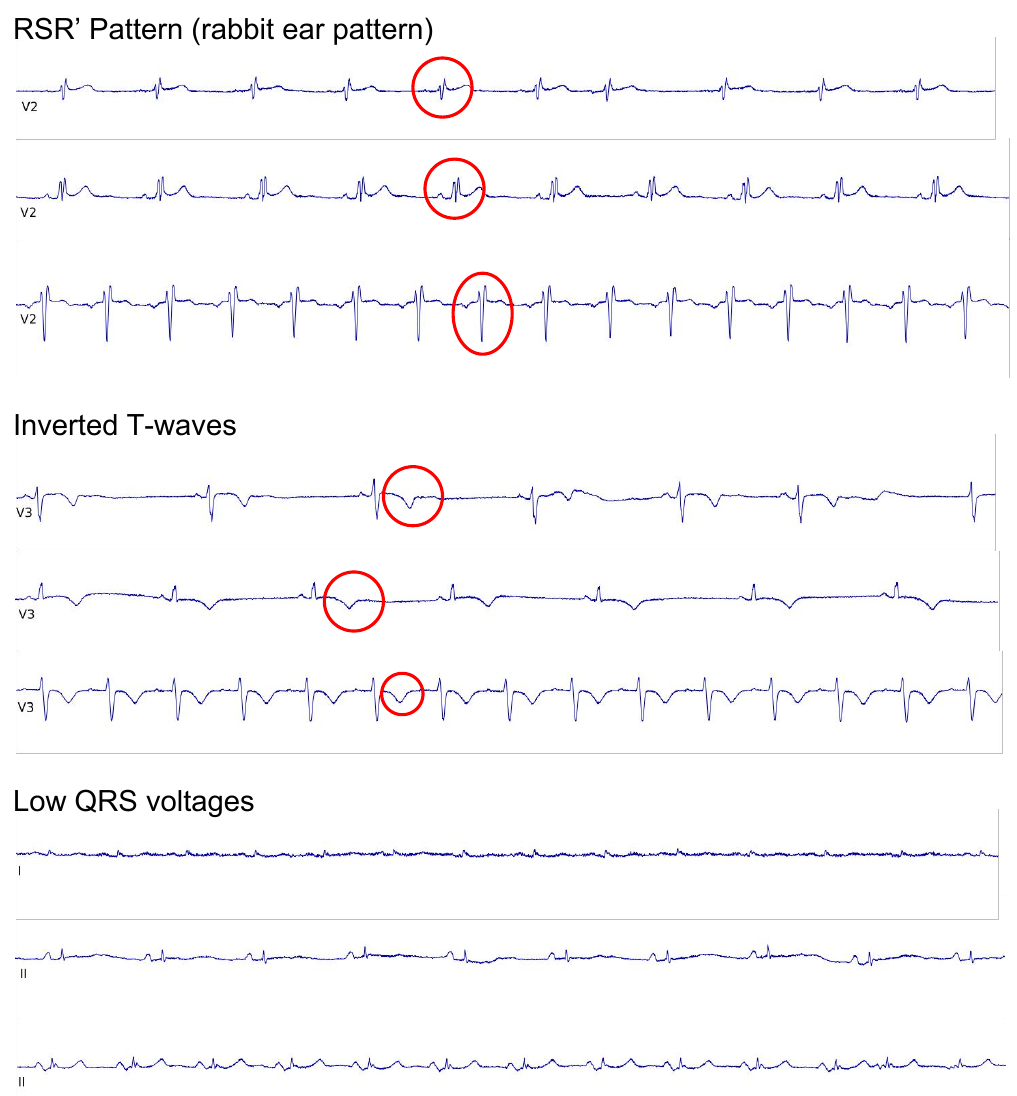}
    \caption{Top 3 retrieved ECG using FOCAL.}
    \label{fig:retrieval}
\end{figure*}

We attempted to use FOCAL to retrieve electrocardiograms (ECGs) from the MIMIC-ECG dataset \citep{gowmimic} through text. To test our model's ability to capture fine-grained waveform features, we tested a series of typical waveform features such as `RSR' Pattern, `Inverted T-waves,' and `Low QRS voltages.' Figure \ref{fig:retrieval} shows the Top 3 retrieved ECGs with probabilities all greater than 0.99. Our model demonstrated strong capability in retrieving ECGs through waveform feature text, which can lead to two applications: (1) Helping doctors quickly find similar cases or specific ECG patterns, aiding in diagnosis and treatment decision-making; (2) In medical education, text-based retrieval can quickly find typical ECG cases, assisting in teaching and training, thereby improving educational effectiveness.

\end{document}